\documentclass{aastex62}
\usepackage{amsmath}

\newcommand\Hzs{Hz~s$^{-1}$ }
\newcommand\Hzsns{Hz~s$^{-1}$}
\newcommand\totcount{26,631,913 }
\newcommand\filterremovedcount{26,588,893 }
\newcommand\filterremovedpercent{$99.84\%$}
\newcommand\postfiltercount{43,020 }
\newcommand\finalcount{$4\,539$ }
\newcommand\injectedcount{20,000 }
\newcommand\recoveredcount{18,528 }
\newcommand\recoveredpercent{$92.64\%$}
\newcommand\recoverednoRFIpercent{$97.66\%$}
\newcommand\recoverednopartnercount{$414$ }
\newcommand\recoverednopartnerdfdtnullcount{$21$ }
\newcommand\recoveredshouldbeFANDYcount{18,093 }
\newcommand\recoveredareFANDYcount{18,044 }
\newcommand\recoveredareFANDYpercent{$99.73\%$}

\received{Sept. 9, 2020}
\revised{Nov. 6, 2020}
\accepted{Nov. 17, 2020}
\submitjournal{AJ}

\shorttitle{A Search for Technosignatures Around 31 Sun-like Stars with the GBT}
\shortauthors{Margot et al.}

\usepackage{hyperref}
\hypersetup{colorlinks=true,citecolor=blue,linkcolor=blue,urlcolor=blue,breaklinks=true}

\begin{document}

\title{A Search for Technosignatures Around 31 Sun-like Stars with the Green Bank Telescope at 1.15--1.73 GHz}

\correspondingauthor{Jean-Luc Margot}
\email{jlm@epss.ucla.edu}

\author[0000-0001-9798-1797]{Jean-Luc Margot}
\affiliation{Department of Earth, Planetary, and Space Sciences, University of California, Los Angeles, CA 90095, USA}
\affiliation{Department of Physics and Astronomy, University of California, Los Angeles, CA 90095, USA}

\author[0000-0003-4736-4728]{Pavlo Pinchuk} %
\affiliation{Department of Physics and Astronomy, University of California, Los Angeles, CA 90095, USA}

\author{Robert Geil} %
\affiliation{Department of Computer Science, University of California, Los Angeles, CA 90095, USA}

\author{Stephen Alexander} %
\affiliation{Department of Physics and Astronomy, University of California, Los Angeles, CA 90095, USA}

\author{Sparsh Arora} %
\affiliation{Department of Computer Science, University of California, Los Angeles, CA 90095, USA}

\author{Swagata Biswas} %
\affiliation{Department of Mathematics, University of California, Los Angeles, CA 90095, USA}

\author{Jose Cebreros} %
\affiliation{Department of Mathematics, University of California, Los Angeles, CA 90095, USA}

\author[0000-0003-1825-4028]{Sanjana Prabhu Desai} %
\affiliation{Department of Earth, Planetary, and Space Sciences, University of California, Los Angeles, CA 90095, USA}

\author[0000-0002-5194-1615]{Benjamin Duclos} %
\affiliation{Department of Physics and Astronomy, University of California, Los Angeles, CA 90095, USA}

\author{Riley Dunne} %
\affiliation{Department of Physics and Astronomy, University of California, Los Angeles, CA 90095, USA}

\author{Kristy Kwan Lin Fu} %
\affiliation{Department of Physics and Astronomy, University of California, Los Angeles, CA 90095, USA}

\author{Shashwat Goel} %
\affiliation{Department of Computer Science, University of California, Los Angeles, CA 90095, USA}

\author{Julia Gonzales} %
\affiliation{Department of Earth, Planetary, and Space Sciences, University of California, Los Angeles, CA 90095, USA}

\author{Alexander Gonzalez} %
\affiliation{Department of Physics and Astronomy, University of California, Los Angeles, CA 90095, USA}

\author{Rishabh Jain} %
\affiliation{Department of Computer Science, University of California, Los Angeles, CA 90095, USA}

\author[0000-0002-4688-314X]{Adrian Lam} %
\affiliation{Department of Electrical Engineering, University of California, Los Angeles, CA 90095, USA}

\author[0000-0002-8984-4319]{Briley Lewis} %
\affiliation{Department of Physics and Astronomy, University of California, Los Angeles, CA 90095, USA}

\author[0000-0001-7003-0588]{Rebecca Lewis} %
\affiliation{Department of Physics and Astronomy, University of California, Los Angeles, CA 90095, USA}

\author{Grace Li} %
\affiliation{Department of English, University of California, Los Angeles, CA 90095, USA}

\author[0000-0003-2562-9043]{Mason MacDougall} %
\affiliation{Department of Physics and Astronomy, University of California, Los Angeles, CA 90095, USA}

\author{Christopher Makarem} %
\affiliation{Department of Computer Science, University of California, Los Angeles, CA 90095, USA}

\author{Ivan Manan} %
\affiliation{Department of Electrical Engineering, University of California, Los Angeles, CA 90095, USA}

\author{Eden Molina} %
\affiliation{Department of Physics and Astronomy, University of California, Los Angeles, CA 90095, USA}

\author{Caroline Nagib} %
\affiliation{Department of Physics and Astronomy, University of California, Los Angeles, CA 90095, USA}

\author{Kyle Neville} %
\affiliation{Department of Mathematics, University of California, Los Angeles, CA 90095, USA}

\author{Connor O'Toole} %
\affiliation{Department of Physics and Astronomy, University of California, Los Angeles, CA 90095, USA}

\author{Valerie Rockwell} %
\affiliation{Department of Physics and Astronomy, University of California, Los Angeles, CA 90095, USA}
\affiliation{Department of Anthropology, University of California, Los Angeles, CA 90095, USA}

\author{Yoichiro Rokushima} %
\affiliation{Department of Mechanical and Aerospace Engineering, University of California, Los Angeles, CA 90095, USA}

\author{Griffin Romanek} %
\affiliation{Department of Computer Science, University of California, Los Angeles, CA 90095, USA}

\author{Carlyn Schmidgall} %
\affiliation{Department of Mathematics, University of California, Los Angeles, CA 90095, USA}
\affiliation{Department of Atmospheric and Oceanic Sciences, University of California, Los Angeles, CA 90095, USA}

\author{Samar Seth} %
\affiliation{Department of Computer Science, University of California, Los Angeles, CA 90095, USA}

\author{Rehan Shah} %
\affiliation{Department of Electrical Engineering, University of California, Los Angeles, CA 90095, USA}

\author[0000-0001-7728-3780]{Yuri Shimane} %
\affiliation{Department of Mechanical and Aerospace Engineering, University of California, Los Angeles, CA 90095, USA}

\author[0000-0002-8793-3246]{Myank Singhal} %
\affiliation{Department of Physics and Astronomy, University of California, Los Angeles, CA 90095, USA}

\author[0000-0002-4675-9069]{Armen Tokadjian}  %
\affiliation{Department of Physics and Astronomy, University of California, Los Angeles, CA 90095, USA}

\author{Lizvette Villafana} %
\affiliation{Department of Physics and Astronomy, University of California, Los Angeles, CA 90095, USA}

\author{Zhixian Wang} %
\affiliation{Department of Computer Science, University of California, Los Angeles, CA 90095, USA}

\author{In Yun} %
\affiliation{Department of Physics and Astronomy, University of California, Los Angeles, CA 90095, USA}

\author{Lujia Zhu} %
\affiliation{Department of Mathematics, University of California, Los Angeles, CA 90095, USA}

\author[0000-0001-5229-7430]{Ryan S.\ Lynch}
\affiliation{Green Bank Observatory, PO Box 2, Green Bank, WV
  24494, USA}

\begin{abstract}
We conducted a search for technosignatures in April of 2018 and 2019
with the L-band receiver (1.15 -- 1.73 GHz) of the 100 m diameter
Green Bank Telescope.  These observations focused on regions
surrounding 31 Sun-like stars near the plane of the Galaxy.  We
present the results of our search for narrowband signals in this data
set as well as improvements to our data processing pipeline.
Specifically, we applied an improved candidate signal detection
procedure that relies on the topographic prominence of the signal
power, which nearly doubles the signal detection count of some
previously analyzed data sets.  We also improved the
direction-of-origin filters that remove most radio frequency
interference (RFI) to ensure that they uniquely link signals observed
in separate scans.  We performed a preliminary signal injection and
recovery analysis to test the performance of our pipeline.  We found
that our pipeline recovers $93\%$ of the injected signals over the
usable frequency range of the receiver and $98\%$ if we exclude
regions with dense RFI.  In this analysis, 99.73\% of the recovered
signals were correctly classified as technosignature candidates.  Our
improved data processing pipeline classified over 99.84\% of the
$\sim$26 million signals detected in our data as RFI.  Of the
remaining candidates, 4539 were detected outside of known RFI
frequency regions.  The remaining candidates were visually inspected
and verified to be of anthropogenic nature.  Our search compares
favorably to other recent searches in terms of end-to-end sensitivity,
frequency drift rate coverage, and signal detection count per unit
bandwidth per unit integration time.
\end{abstract}


\vfill
\section{Introduction} \label{sec:intro}

We describe a search for radio technosignatures with the
L-band receiver of the 100 m diameter Green Bank Telescope (GBT).  We
used 
a total of 4 hours of GBT time in 2018 and 2019 to observe the regions
around 31 Sun-like stars near the plane of the Galaxy.  We have so far
prioritized the detection of narrowband ($\sim10$ Hz) signals because
they are diagnostic of engineered emitters \citep[e.g.,][]{Tarter2001}.

Our search builds on the legacy of technosignature searches performed
in the period 1960--2010 \citep[][and references therein]{Tarter2001,
  Tarter2010} and previous searches conducted by our group
\citep[][]{Margot2018, Pinchuk2019}.  Other recent searches include
work conducted by \citet{Siemion2013, Harp2016, Enriquez2017, Gray2017,
  Price2020}.

Our choice of search parameters has key advantages compared to the
Breakthrough Listen (BL) searches described by \citet{Enriquez2017}
and \citet{Price2020},
which contend with much larger data volumes.  Specifically, our search provides roughly
uniform detection sensitivity over the entire range of frequency drift
rates ($\pm8.86$ \Hzsns) whereas the BL searches suffer a substantial
loss in sensitivity due to the spreading of signal power across up to
13--26 frequency resolution cells.  In addition, we cover a range of
frequency drift rates that is 2--4 wider than the BL searches with a
time resolution that is 51 times
better.

Our search algorithms
are distinct from the BL algorithms in that they alleviate the
necessity of discarding $\sim$kHz wide regions of frequency space
around every detected signal.  We abandoned this practice in previous
work \citep{Pinchuk2019}.  In this work, we further refined our
algorithm by implementing a candidate signal detection procedure that
relies on the concept of prominence and by removing the requirement to
compute the bandwidth of candidate signals.
Our new approach, combined with better end-to-end sensitivity and
drift rate coverage, enables a hit rate density or signal detection
count per unit bandwidth per unit integration time that is $\sim$\replaced{300}{200} times
larger than that of the BL search described by \citet{Price2020}.

A key measure of the robustness and efficiency of a data processing
pipeline is provided by the technique of signal injection and recovery
\citep[e.g.,][]{Christiansen2013}, whereby artificial signals are injected into
the raw data and the fraction of signals recovered by the pipeline is
quantified.  Despite the importance of this metric, we are not aware
of an existing tool to quantify the recovery rates of data-processing
pipelines in radio technosignature searches.  We make a first step
towards the implementation of this tool and show that our current
pipeline detects $93\%$ of the injected signals over
the usable frequency range of the receiver and $98\%$ if we exclude
regions with dense RFI.
In addition, our pipeline correctly flagged \recoveredareFANDYpercent
of the detected signals as technosignature candidates.  Although our
current implementation requires additional work to fully capture the
end-to-end pipeline efficiency, it can already illuminate
imperfections in our and other groups' pipelines and be used to
calibrate claims about the prevalence of other civilizations
\citep[e.g.,][]{Enriquez2017}.

The article is organized as follows.  Our data acquisition and analysis techniques are presented in Sections \ref{sec:data aq} and \ref{sec:analysis}, respectively.  Our preliminary signal injection and recovery analysis is described in Section \ref{sec:injection_analysis}.
The main results of our search are outlined in Section \ref{sec:results}.
In Section \ref{sec:discussion}, we describe
certain advantages of our search, including
dechirping efficiency, drift rate coverage, data archival practices, 
candidate detection algorithm, and hit rate density.  We also discuss
limits on the prevalence of other civilizations,
search metrics such as the Drake Figure of Merit, and re-analysis of
previous data with our latest algorithms.  We close with our
conclusions in Section \ref{sec:conclusions}.

\section{Data Acquisition and Pre-Processing} \label{sec:data aq}

Our data acquisition techniques are generally similar to those used by \citet{Margot2018} and \citet{Pinchuk2019}.  Here, we give a brief overview and refer the reader to these other works for additional details.

\subsection{Observations} \label{subsec:sources}

We selected 31 Sun-like stars (spectral type G, luminosity class V)
with median galactic latitude of 0.85$^\circ$ (Table \ref{tab:soucres})
because their properties are similar to the only star currently 
known to harbor a planet with
life.  We observed these stars with the GBT during two 2-hour sessions
separated by approximately one year.  During each observing session,
we recorded both linear polarizations of the L-band receiver with the
GUPPI backend in its baseband recording mode \citep{GUPPI}, which
yields 2-bit raw voltage data after requantization with an optimal
four-level sampler \citep{Kogan1998}.  The center frequency was set to
1.5 GHz and we sampled 800 MHz of bandwidth between 1.1 and 1.9~GHz,
which GUPPI channelized into 256 channels of 3.125 MHz each.  We
validated the data acquisition and analysis processes at the beginning
of each observing session by injecting a monochromatic tone near the
receiver frontend and recovering it at the expected frequency
in the processed data.

We observed all our targets in pairs in order to facilitate the
detection and removal of signals of terrestrial origin (Section
\ref{subsec:SQL_filters}).  The sources were paired in a way that
approximately minimized telescope time overhead, i.e., the sum of the
times spent repositioning the telescope.  Pairings were adjusted to
avoid pair members that were too close to one another on the plane of
the sky 
with the goal of 
eliminating any possible ambiguity in the direction of
origin of detected signals. Specifically, we required angular
separations larger than $1^{\circ}$ between pair members, i.e.,
several times the $\sim8.4$ arcmin beamwidth of the GBT at 1.5 GHz.

Each pair was observed twice in a 4-scan sequence: A, B, A, B.  The
integration time for each scan was 150~s, yielding a total integration
time of 5 minutes per target.  CoRoT 102810550 and CoRoT 110777727
were each observed for an additional two scans.
With 66 scans of 150 s duration each, our total integration  time amounts to 2.75 hr.

\begin{deluxetable}{lcCCCCc}[b]
  \caption{Target host stars listed in order of observation.
    Successive pairs are separated by a blank line. Spectral types, galactic coordinates, and parallax measurements were obtained from the SIMBAD database \citep{Wenger2000}.
    Distances in light years (ly) were calculated from the parallax measurements.
    The Modified Julian Date (MJD) refers to the beginning of the first scan.
    \label{tab:soucres}}
\tablehead{
  \colhead{Host Star} & \colhead{Spectral Type} & \colhead{Long. (deg)} & \colhead{Lat. (deg)} & \colhead{Parallax (mas)} & \colhead{Distance (ly)} & \colhead{MJD of Scan 1}}  

\startdata
\multicolumn{7}{c}{April 27, 2018 20:00 -- 22:00 UT}\\
\hline
TYC 1863-858-1        & G0V & $185.5$ & $-0.21$ & $1.9547 \pm 0.036  $ & $1669 \pm 31$  & 58235.84503472    \\       
TYC 1868-281-1        & G2V & $185.3$ & $-0.65$ & $3.8622 \pm 0.046  $ & $844 \pm 10$   & 58235.84737269    \\ [1ex] 
HD 249936             & G2V & $186.2$ & $-0.80$ & $1.9515 \pm 0.0412 $ & $1671 \pm 35$  & 58235.85430556    \\ 
TYC 1864-1748-1       & G2V & $186.6$ & $+0.87$ & $3.0350 \pm 0.043  $ & $1075 \pm 15$  & 58235.85671296    \\ [1ex]
HIP 28216             & G2V & $186.9$ & $-0.93$ & $1.2479 \pm 0.0909 $ & $2614 \pm 190$ & 58235.86393519    \\ 
HD 252080             & G5V & $188.1$ & $+1.18$ & $5.7621 \pm 0.0511 $ & $566 \pm 5$    & 58235.86640046    \\ [1ex]
HD 251551             & G2V & $186.5$ & $+1.65$ & $4.5105 \pm 0.0755 $ & $723 \pm 12$   & 58235.87356481    \\ 
HD 252993             & G0V & $186.3$ & $+3.13$ & $6.9544 \pm 0.0401 $ & $469 \pm 3$    & 58235.87591435    \\ [1ex]
TYC 742-1679-1        & G5V & $195.8$ & $-2.07$ & $8.4009 \pm 0.0360 $ & $388 \pm 2$    & 58235.88324074    \\ 
HD 255705             & G5V & $196.9$ & $-0.04$ & $6.8001 \pm 0.0570 $ & $480 \pm 4$    & 58235.88562500    \\ [1ex]
HD 254085             & G0V & $197.5$ & $-1.96$ & $6.5338 \pm 0.0967 $ & $499 \pm 7$    & 58235.89270833    \\ 
HD 256380             & G8V & $198.0$ & $-0.02$ & $2.3058 \pm 0.0398 $ & $1415 \pm 24$  & 58235.89505787    \\ [1ex]
TYC 739-1501-1        & G2V & $198.2$ & $-1.60$ & $ ... $              & $ ... $        & 58235.90204861    \\ 
HD 256736             & G2V & $198.2$ & $+0.22$ & $6.1808 \pm 0.0802 $ & $528 \pm 7$    & 58235.90435185    \\ [1ex]
TYC 739-1210-1$^{a}$  & G5V & $198.5$ & $-1.16$ & $9.9809 \pm 0.0424 $ & $327 \pm 1$    & 58235.91119213    \\ \hline
\multicolumn{7}{c}{April 26, 2019 22:00 -- 24:00 UT}\\
\hline
TYC 148-515-1    & G5V & $212.4 $ & $-0.98$ & $5.4947 \pm 0.0416$ & $594 \pm 4$     &  58599.92175926    \\
CoRoT 102810550  & G2V & $211.5 $ & $-0.69$ & $1.1333 \pm 0.0247$ & $2878 \pm 63$   &  58599.92392361    \\[1ex]
CoRoT 102830606  & G2V & $211.4 $ & $-0.49$ & $2.2193 \pm 0.0357$ & $1470 \pm 24$   &  58599.93030093    \\
TYC 149-362-1    & G5V & $212.8 $ & $+0.69$ & $1.1606 \pm 0.0581$ & $2810 \pm 141$  &  58599.93254630    \\[1ex]
TYC 149-532-1    & G2V & $213.1 $ & $+0.69$ & $7.2260 \pm 0.0380$ & $451 \pm 2$     &  58599.93917824    \\
CoRoT 102827664  & G4V & $211.4 $ & $-0.51$ & $2.3394 \pm 0.0274$ & $1394 \pm 16$   &  58599.94144676    \\[1ex]
CoRoT 102936925  & G4V & $213.6 $ & $-0.93$ & $1.0454 \pm 0.0223$ & $3120 \pm 67$   &  58599.94826389    \\
CoRoT 110695685  & G4V & $215.9 $ & $-0.83$ & $1.5743 \pm 0.0541$ & $2072 \pm 71$   &  58599.95049769    \\[1ex]
CoRoT 110864307  & G2V & $216.1 $ & $-0.98$ & $1.5985 \pm 0.0442$ & $2040 \pm 56$   &  58599.95706019    \\
CoRoT 102951397  & G2V & $213.6 $ & $-0.87$ & $1.1909 \pm 0.0249$ & $2739 \pm 57$   &  58599.95931713    \\[1ex]
CoRoT 102963038  & G3V & $213.7 $ & $-0.85$ & $0.3134 \pm 0.0256$ & $10407 \pm 850$ &  58599.96596065    \\
HD 50388         & G8V & $215.2 $ & $-0.75$ & $7.3465 \pm 0.0598$ & $444 \pm 4$     &  58599.96820602    \\[1ex]
TYC 4805-3328-1  & G5V & $215.4 $ & $-0.19$ & $2.5383 \pm 0.0455$ & $1285 \pm 23$   &  58599.97480324    \\
CoRoT 110777727  & G1V & $216.2 $ & $-0.90$ & $1.5407 \pm 0.0457$ & $2117 \pm 63$   &  58599.97699074    \\[1ex]
CoRoT 110776963  & G4V & $215.9 $ & $-0.78$ & $2.5207 \pm 0.0436$ & $1294 \pm 22$   &  58599.98373843    \\
TYC 4814-248-1   & G2V & $215.1 $ & $+1.32$ & $2.9668 \pm 0.0426$ & $1099 \pm 16$   &  58599.98597222    \\[1ex]
\enddata
\tablenotetext{a}{The source paired with TYC 739-1210-1 was observed only once and not analyzed.}
\end{deluxetable}

\subsection{Sensitivity}
\citet{Margot2018} calculated the sensitivity of a search for
narrowband signals performed with the 100 m GBT.  Assuming a System
Equivalent Flux Density (SEFD) of 10 Jy, integration time of 150 s,
and frequency resolution of 3~Hz, they found that sources with flux
densities of 10 Jy can be detected with a signal-to-noise ratio (S/N)
of 10.  The results of that calculation are directly applicable here
because our search parameters are identical to that study.
Specifically, our search is sensitive to \replaced{an Arecibo
  planetary radar transmitter}{ transmitters with the effective
  isotropic radiated power (EIRP) of the Arecibo planetary radar
  transmitter (2.2 $\times 10^{13}$ W)} located 420 ly from Earth
\citep[][Figure 5]{Margot2018}.  Transmitters located as far as the
most distant source (CoRoT 102963038; $\sim10,407$ ly) and with
$<$1000 times the Arecibo \replaced{effective isotropic radiated power
  (EIRP)}{EIRP} can also be detected in this search. Although we
selected Sun-like stars as primary targets, our search is obviously
sensitive to other emitters located within the beam of the telescope.
A search of the Gaia DR2 catalog \citep{gaiadr1,gaiadr2} inspired by
\citet{wlod20} reveals that there are \replaced{15,013}{15,031} known
stars with measured parallaxes within the half-power beamwidths
associated with our 31 primary sources.  \added{The median and mean
  distances to these sources are 2088 and 7197 ly, respectively}.

\subsection{Computation of Power Spectra} \label{subsec:preproc}

After unpacking the
digitized raw voltages from 2-bit to 4-byte floating point values, we
computed consecutive power spectra with $2^{20}$-point Fourier
transforms, yielding a frequency resolution of $\Delta f = 2.98$ Hz.
We chose this frequency resolution because it is small enough to
provide unambiguous detections of narrowband ($<$10 Hz)
technosignatures and large enough to examine Doppler frequency drift
rates of up to nearly $\pm10$ Hz~s$^{-1}$ (Section
\ref{subsec:treealg}).  We processed all channels within the operating
range of the GBT L-band receiver 
(1.15--1.73 GHz), 
excluding channels
that overlap the frequency range (1200--1341.2 MHz) of a notch filter
designed to mitigate radio frequency interference (RFI) from a nearby
aircraft detection radar\replaced{.}{, for a total processed bandwidth of 438.8 MHz.}
Although \citet{Enriquez2017} and \citet{Price2020} used the L-band
receiver over a larger frequency range (1.1--1.9 GHz), we used a
narrower range because we observed serious degradation of the bandpass
response beyond the nominal operating range of the receiver.

In order to correct for the bandpass response of GUPPI's 256 channels,
we fit a 16-degree Chebyshev polynomial to the median bandpass
response of a subset of the processed channels that did not include
strong RFI and that were not close to the cutoff frequencies of
filters located upstream of the GUPPI backend.
After applying the bandpass correction to all channels, we stored the
consecutive power spectra as rows in time-frequency arrays (a.k.a
time–frequency diagrams, spectrograms, spectral waterfalls, waterfall
plots, or dynamic spectra) and normalized the power to zero mean and
unit standard deviation of the noise power.  The normalized power
values reflect the
S/N
at each time and frequency bin.

\subsection{Doppler Dechirping} \label{subsec:treealg}

Due to the orbital and rotational motions of both the emitter and the
receiver, we expect extraterrestrial technosignatures to drift in
frequency space \citep[e.g.,][]{Siemion2013, Margot2018, Pinchuk2019}.
To integrate the signal power over the scan duration while
compensating for Doppler drifts in signal frequency, we used
incoherent sums of power spectra, where each individual spectrum was
shifted in frequency space by a judicious amount prior to summation.
This technique is known as {\em incoherent dechirping}.
Coherent dechirping algorithms exist \citep{korp12} but
are computationally expensive and seldom used.

Because the Doppler drift rates due to the emitters are unknown, we
examined 1023 linearly spaced drift rates in increments of $\Delta
\dot{f} = 0.0173$~\Hzsns over the range $\pm8.86$ \Hzsns.  To
accomplish this task, we used a computationally advantageous
tree algorithm \citep{Taylor1974, Siemion2013}, which
operates on the dynamic spectra and yields time integrations of the
consecutive power spectra after correcting approximately for each
trial Doppler drift rate.
The algorithm requires input spectra with a number of rows equal to a
power of two, and we zero-padded the dynamic spectra with
approximately 65 rows to obtain 512 rows.
The output of this algorithm, 
which was run once for negative drift rates and once for positive drift rates,
is stored in
1023 $\times$ 2$^{20}$ drift-rate-by-frequency arrays that are ideal
for identifying {\em candidate signals}, i.e., radio signals that
exceed a certain detection threshold (Section
\ref{subsec:candidate_signal_detection}).  We quantify the sensitivity
penalty associated with the use of the tree algorithm in
Section~\ref{subsec:drift_rates_comparison}.

\section{Data Analysis} \label{sec:analysis}

\subsection{Candidate Signal Detection} \label{subsec:candidate_signal_detection}

We performed an iterative search for candidate signals on the
drift-rate-by-frequency arrays obtained with the 
incoherent dechirping algorithm.  Specifically, we identified the
signal with the highest integrated S/N and stored its characteristics
in a structured query language (SQL) database, then identified and
recorded the signal with the second highest S/N, and so on.  Redundant
detections can occur when signals in the vicinity of a candidate
signal have large integrated power along similar drift
rates. Different data processing pipelines tackle these redundant
detections in different ways.  \citet{Siemion2013},
\citet{Enriquez2017}, \citet{Margot2018} and \citet{Price2020}
discarded all detections within $\sim$kHz-wide regions of frequency
space around every candidate signal detection.  This method leaves
large portions of the observed frequency space unexamined and biases
the results towards high S/N signals because signals with lower S/N in
their vicinity are discarded. Importantly, this method
complicates 
attempts to place upper limits on the abundance of technosignature
sources
because the pipeline eliminates the very signals it purports to detect
(Sections \ref{subsec:cand_detection_comparison} and
\ref{subsec:existence_limits_discussion}).

\begin{figure}[ht!]
\plotone{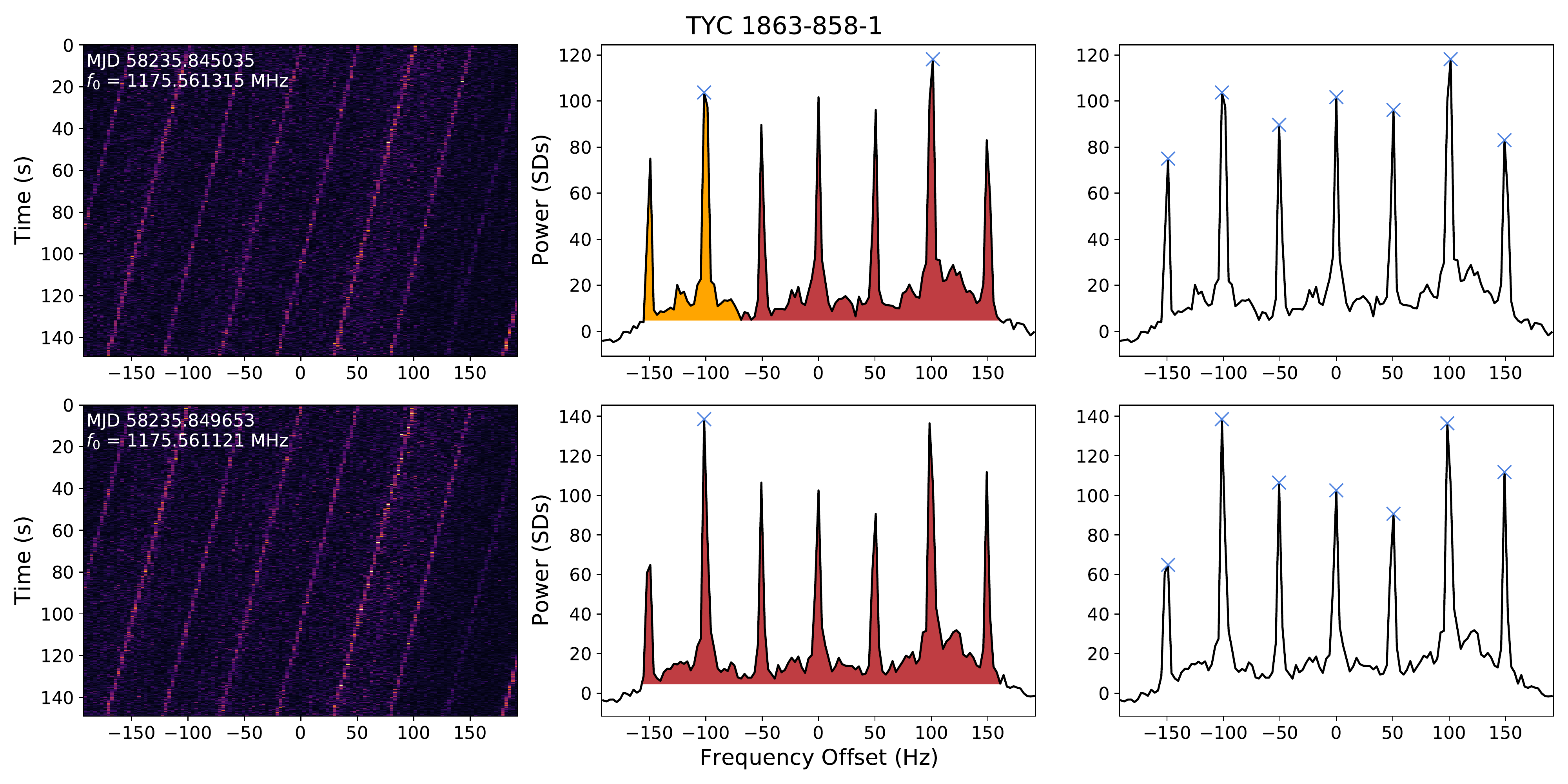}
\caption{Comparison of signal detection procedures illustrated on a $\sim$400 Hz region for scans 1 (top) and 2 (bottom) of TYC~1863-858-1. 
  (Left)
  Dynamic spectra,
  where pixel intensity represents signal power.
  (Middle) Integrated power spectra with blue crosses marking the signals that
  are detected with the procedure described by
  \citet{Pinchuk2019}.  In the first scan, the strongest signal (+100 Hz) is detected and the corresponding 5$\sigma$ bandwidth is shown in red.
  The second strongest signal ($-$100 Hz) is then detected and the corresponding 5$\sigma$ bandwidth is shown in orange.
  In the second scan, only the strongest signal, which is now at $-100$Hz, is detected.
(Right) Integrated power spectra with blue crosses marking the signals that are detected with the procedure described in this work.
  \label{fig:CDBad}}
\end{figure}

\citet{Pinchuk2019} introduced a novel procedure to alleviate these
shortcomings.  In order to avoid redundant detections, they imposed
the restriction that two signals cannot cross in in the time-frequency
domain of the scan and discarded detections only in a small frequency
region around every candidate signal detection.  The extent of this
region was set equal to the bandwidth of the candidate signal measured
at the $5\sigma$ power level, where $\sigma$ is one standard deviation
of the noise.  Unfortunately, this bandwidth calculation can cause
complications in some situations.
For example, small noise fluctuations may result in an unequal number
of candidate signals detected in two scans of a source.  In the
$\sim$400 Hz region of the spectrum shown on Figure \ref{fig:CDBad},
two signals ($\pm 100$Hz) are detected in the first scan but only one
signal ($- 100$Hz) is detected in the second scan.  This
incompleteness is detrimental to our direction-of-origin filters
(Section \ref{subsec:SQL_filters}), which rely on accurate signal
detection across all scans. Moreover, discarding the region
corresponding to a bandwidth measured at $5\sigma$
prevents the detection of at least five other signals per scan 
in this example (Figure \ref{fig:CDBad}).

In this work, we improve on the procedure presented by
\citet{Pinchuk2019} in two important ways.  First, we identify
candidate signal detections on the basis of the topography-inspired
concept of {\em prominence}.  The prominence of a signal is defined as
the vertical distance between the peak and its lowest contour line, as
implemented in the numerical computing package \textit{SciPy}
\citep{SciPy}.  Because our integrated spectra are one-dimensional, we
take the larger of a peak's two `bases' as a replacement for the
lowest contour line.
The high-frequency (low-frequency) base is defined as the minimum
power in the frequency region starting on the high (low) frequency
side of the peak and ending +500 Hz ($-$500 Hz) away or at the frequency
location of the nearest peak with higher (lower) frequency and larger
power, whichever results in the smallest frequency interval.
While the $\pm$500 Hz limits are not essential to compute prominences,
they do speed up the calculations.
Second, we remove the bandwidth-dependence of \citet{Pinchuk2019}'s algorithm.
Instead, we apply a local maximum filter to the
drift-rate-by-frequency arrays in order to remove any points that are
not a maximum in their local 3x3 neighborhood.  We find that this
filter in conjunction with the prominence-based candidate signal
detection identifies the signals of interest without introducing
redundant detections.

\begin{figure}[ht!]
  \begin{center}
    \includegraphics[width=4in]{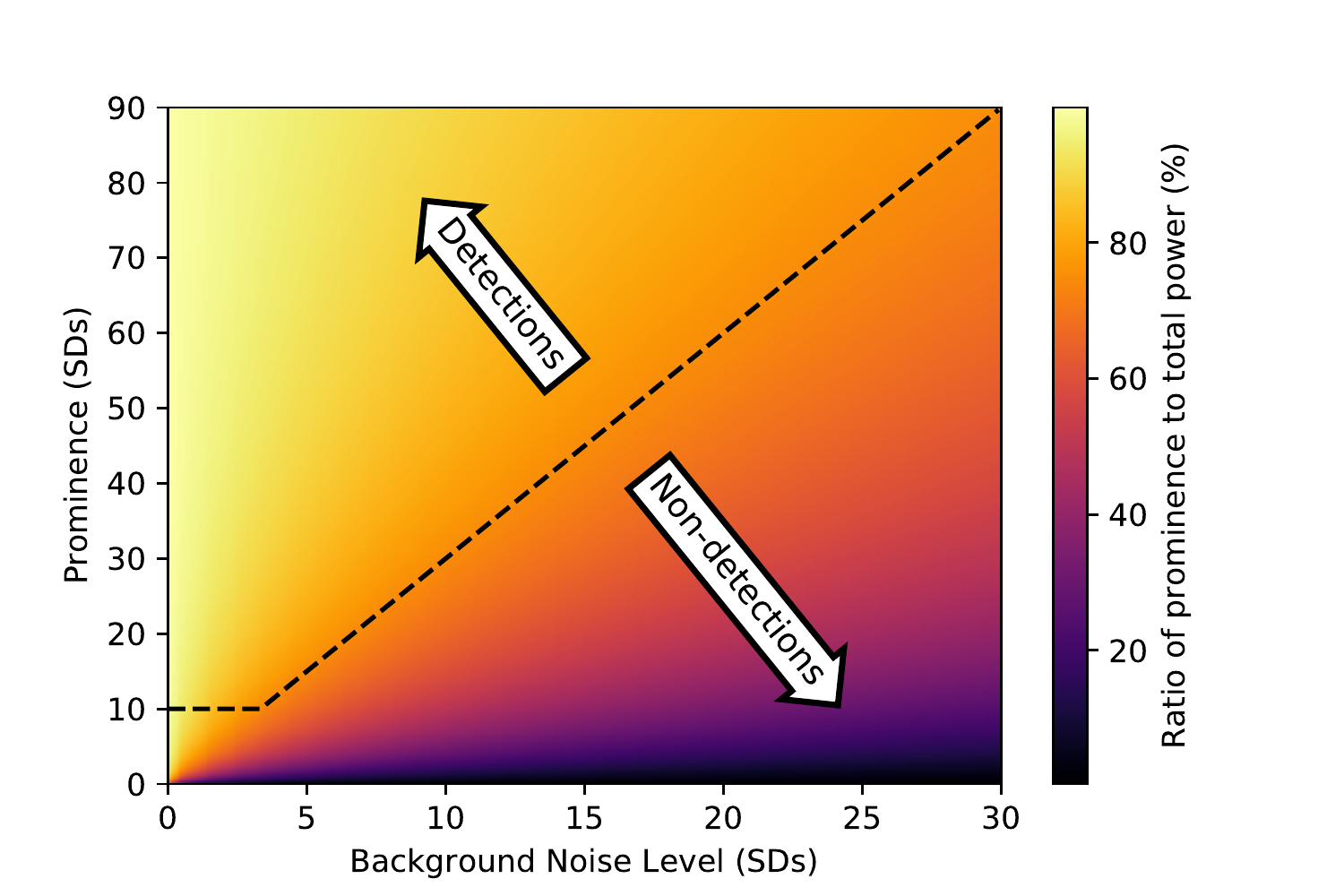}
  \end{center}
  \caption{
  Illustration of detection criteria.  Signals above the dashed black line line are marked as detections by our pipeline.  
   \label{fig:detection_space}}       
\end{figure}

Signals are considered candidate detections if their prominence meets
two criteria: (1) their prominence exceeds $10\sigma$, where $\sigma$
is one standard deviation of the noise in the integrated spectrum,
(2) their prominence
exceeds a fraction $f$ of their integrated power.
For this analysis, we settled on $f=75\%$.
The second requirement is necessary because power fluctuations
superimposed on strong broadband signals that approach or exceed the
$10\sigma$ detection threshold can yield prominences that exceed
$10\sigma$.
With this second requirement,
a signal with a prominence of $10\sigma$ above a 3.0-$\sigma$ baseline
would be marked as a detection, but the same signal above a
3.5-$\sigma$ baseline would not.  
Figure \ref{fig:detection_space} describes the detection space.

\begin{figure}[ht!]
  \begin{center}
    \includegraphics[width=4in]{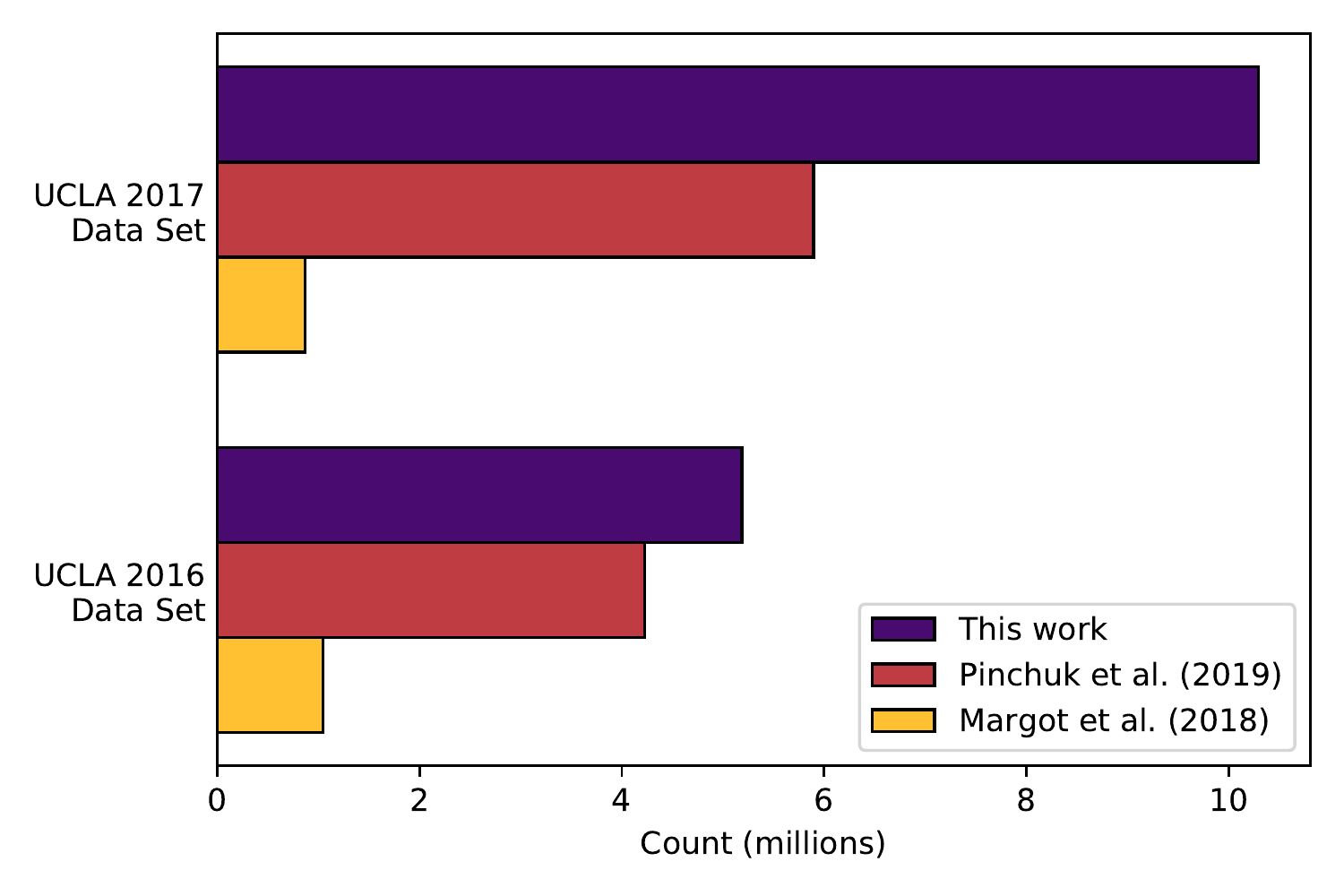}
  \end{center}
\caption{
  Detection counts obtained with the algorithms presented by \citet{Margot2018}, \citet{Pinchuk2019}, and this work.
  Our current pipeline detects 1.23--1.75 as many signals as \citet{Pinchuk2019}'s pipeline and $\sim 12$ times as many signals as \citet{Margot2018}'s pipeline.  
   \label{fig:signal_counts_detection}}       
\end{figure}

As a result of these candidate signal detection improvements, we now detect 
1.23--1.75 and $\sim$12 times as many signals as we did with the data
processing pipelines of \citet{Pinchuk2019} and \citet{Margot2018},
respectively (Figure \ref{fig:signal_counts_detection}).
We compare this signal detection performance to that of other searches
in Section \ref{subsec:cand_detection_comparison}.

Once a signal with frequency $f_0$ and drift rate $\dot f_0$ is
detected with the criteria described in this section, 
we follow the procedure outlined by \citet{Pinchuk2019}.
Specifically, we eliminate any other candidate signal with frequency
$f$ and drift rate $\dot f$ if the following inequalities
hold true at the start of the scan:
\begin{eqnarray} \label{eq:blanking}
f_0  <  f <  f_0 + (\dot f - \dot f_0 ) \tau  \qquad \text{for} \quad \dot f > \dot f_0 \\
\nonumber
f_0  >  f >  f_0 + (\dot f - \dot f_0 ) \tau  \qquad \text{for} \quad \dot f < \dot f_0
\end{eqnarray}
where $\tau$ is the scan duration.  Our candidate detection procedure
was applied iteratively until all candidate signals with
prominence $\geq 10\sigma$ were identified.
Occasionally, signals with prominences $\geq 10\sigma$ but S/N $<$ 10
get recorded in the database.  This condition tends to occur primarily
in
regions with dense RFI where the baseline subtraction is imperfect.
For this reason, we flagged all signals with S/N $<$ 10 and did not
consider them to be valid candidates.

\subsection{Doppler and Direction-of-Origin Filters} \label{subsec:SQL_filters}

After identifying all candidate signals, we applied a Doppler filter
and improved variants of our direction-of-origin filters
\citep{Margot2018, Pinchuk2019} to detect and discard anthropogenic
signals in the data.

We began by applying a Doppler filter, which is designed to remove all
signals with zero Doppler drift rate, defined here as signals that
drift less than one frequency resolution cell ($\Delta f$ = 2.98~Hz)
over the duration of a scan ($\tau$ = 150~s).
The signals removed by this filter are of no interest to us
because the corresponding emitters exhibit no line-of-sight
acceleration with respect to the receiver, suggesting that they are
terrestrial in nature.

Next, we applied two direction-of-origin filters, which are designed
to remove any signal that is either not persistent (i.e., not detected
in both scans of its source) or detected in multiple directions on the
sky (i.e., also detected in scans corresponding to other sources).
Because the largest possible sidelobe gain is approximately -30 dB
compared to the main lobe gain, signals detected in multiple
directions on the sky are almost certainly detected through antenna
sidelobes.  The second filter is highly
effective at removing such signals.

As explained by \citet{Pinchuk2019}, the direction-of-origin filters
compare signals from different scans and flag them according to the
observed relationships.  For example, if a signal from a scan of
source A is paired with a signal from a scan of source B, then both
signals are removed because they are detected in multiple directions
of the sky.  In our previous implementation of these filters, two
signals were considered a pair if their drift rates were similar and
their frequencies at the beginning of each scan were within a
predetermined tolerance of a straight line with a slope corresponding
to the drift rate.  With this definition, it was possible for
\textit{multiple} signals in one scan to be paired with a
\textit{single} signal from a different scan, which is undesirable.
For example, a valid technosignature candidate from one of the scans
of source `A' could be labeled as RFI because it was paired with a
signal from one of the scans of source `B', even if the signal in the
scan of source `B' was already paired with a different (RFI) signal
from the scan of source `A'.

We have redesigned our filter implementation to keep a record of all
signals that are paired during filter execution.  We use this record
to impose the restriction that each signal is allowed to pair with
only one other signal in each scan.
Additionally, we implemented an
improved pairing procedure that is loosely based on the Gale – Shapley
algorithm \citep{Gale1962} designed to solve the stable matching
problem.  Our improved procedure operates as follows.  We define the
propagated frequency difference $\Delta F(f_i, f_j)$ of two signals
from different scans to be
\begin{equation}\label{eq:PropFreqDiff}
\Delta F(f_i, f_j) = \left|(f_i - f_j) + \overline{\dot f_{ij}} \Delta t_{ij}\right|
\end{equation}
where $f_i$ and $f_j$ are the start frequencies of the two signals,
$\overline{\dot f_{ij}} = (\dot f_i + \dot f_j) / 2$ is the average of
the two signal drift rates, and $\Delta t_{ij} = t_j - t_i$ is the
time difference between the two scans.  Our updated algorithm iterates
over all remaining unpaired candidate signals and updates the pairings
until $\Delta F(f_i, f_j)$ is minimized for all signal pairs.  In rare
cases when the minimum value of $\Delta F(f_i, f_j)$ is not unique,
multiple pairings are allowed, but no inference about the
anthropogenic nature of the signals is made on the basis of these
pairings alone.

To ensure that paired signals likely originated from the same emitter,
we impose two additional requirements on all signal pairs.  First, we
require that
\begin{equation}\label{eq:FreqReq}
f_{ij, -} \leq f_j \leq f_{ij, +} \; \; {\rm or} \; \; f_{ji, -} \leq f_i \leq f_{ji, +}
\end{equation}
where $f_{ij, \pm} = f_i + (\dot f_i \pm \Delta \dot{f})\Delta t_{ij}  \pm \Delta f$ 
represent the propagated frequency bounds and 
$\Delta \dot{f}$, $\Delta f$ are the drift rate and frequency 
resolution, given by 0.0173 \Hzsns and 2.98 Hz, respectively. 
This condition places an upper limit on $\Delta F(f_i, f_j)$
and we reject signal pairs whose 
propagated frequency differences exceed this bound.
Second, we require that 
\begin{equation}\label{eq:DOOcrit2}
\left| \dot f_i - \dot f_j \right| \leq 2 \Delta \dot{f},
\end{equation}
and we reject signal pairs that do not satisfy this criterion.
In tandem, these requirements reduce the possibility of pairing 
two unrelated signals.

To determine if a signal is persistent (first filter), we apply the
pairing procedure to candidate signals detected in both scans of a
source.  Those signals left without a partner are deemed to originate
from transient sources and are labeled as RFI.  To determine whether a
signal is detected in multiple directions of the sky (second filter),
we apply the pairing procedure to signals from scans of different
sources.  In this case, all resulting pairs are attributed to RFI and
discarded.  Candidate signals remaining after the application of
these procedures
are marked for further inspection.

\subsection{Frequency Filters} \label{subsec:Freq_filters}

A majority of the candidate signals detected in our search are found
in operating bands of known interferers.  Table \ref{tab:known_RFI}
describes the frequency ranges and signal counts associated with the
most prominent anthropogenic RFI detected in our data.  Candidate
signals detected within these frequency regions \added{(except the ARSR products region)} were removed from
consideration because of their likely anthropogenic nature.
\added{The combined 2017 and 2018 signal detection counts in the excluded RFI regions (156,327/MHz) are
  considerably higher than outside of these regions (20,654/MHz) or in
  the 1400--1427 MHz radio astronomy protected band (6,949/MHz).  The
  protected band is regrettably polluted, possibly as a result of
  intermodulation products generated at the telescope \citep{Margot2018}.}

The useful bandwidth of our observations $\Delta f_{\rm tot} = 309.3$
MHz is computed by taking the operational bandwidth of the GBT L-band
receiver (580 MHz) and subtracting the bandwidth of the GBT notch
filter (141.2 MHz) and the total bandwidth discarded due to known
interferers (Table \ref{tab:known_RFI}; 129.5 MHz).

\begin{deluxetable}{lrrrl}[h!]
  \caption{Definitions of operating regions of known anthropogenic interferers and associated signal counts.
The column labeled ``Post-filter Count'' lists the number of signals remaining after application of our Doppler and direction-of-origin filters.
The time-frequency structure of the RFI labeled as ``ARSR products'' is similar to that described by \citet{Siemion2013}, \citet{Margot2018}, and \citet{Pinchuk2019}.  These products are likely intermodulation products of Air Route Surveillance Radars (ARSR).
\label{tab:known_RFI}}
\tablehead{
\colhead{Frequency Region (MHz)} & \colhead{Total Detection Count} & \colhead{\% of Total Detections} & \colhead{Post-filter Count} & \colhead{Identification}}
\startdata
1155.99 -- 1196.91 &  11,937,074  &  44.82 \%  &  15,034   &  GPS L5               \\
1192.02 -- 1212.48 &  135,769      &  0.51  \%  &  276       &  GLONASS L3           \\
1422.32 -- 1429.99 &  190,530      &  0.72  \%  &  2\,945    &  ARSR products       \\
1525 -- 1559       &  8,258,612   &  31.01 \%  &  341       &  Satellite downlinks  \\
1554.96 -- 1595.88 &  5,016,951   &  18.84 \%  &  19,621   &  GPS L1               \\
1592.95 -- 1610.48 &  933,813      &  3.51  \%  &  3,569    &  GLONASS L1           \\
\enddata
\end{deluxetable}

\section{Preliminary Signal Injection and Recovery Analysis} \label{sec:injection_analysis}

A signal injection and recovery analysis consists of injecting
artificial signals into the raw data and quantifying the fraction of
signals that are properly recovered by the pipeline \citep[e.g.,][]{Christiansen2013}.
Although a rigorous injection analysis is beyond the scope of this
paper, we performed a preliminary examination by injecting narrowband
(2.98 Hz) signals into the dynamic spectra before applying the
incoherent dechirping (Section \ref{subsec:treealg}), candidate detection
(Section \ref{subsec:candidate_signal_detection}), and Doppler and
direction-of-origin
filtering (Section \ref{subsec:SQL_filters}) procedures.

\subsection{Generation and Injection of Artificial Signals} \label{subsec:injection_candidates}

We selected 10\,000 starting frequencies from a uniform distribution
over the operating region of the GBT L-band receiver (1.15--1.73 GHz),
excluding the frequency region of the GBT notch filter (1.2--1.3412 GHz).
For each starting frequency, we also randomly selected a frequency
drift rate from the discrete set 
$\{ k \times \Delta \dot{f} \, : \, k \in \mathbb{Z},\, -510 \leq k \leq 510 \}$,
with $\Delta \dot{f}$ = 0.0173 \Hzsns.
Each signal was randomly assigned to one of the sources and was
injected into the first scan of this source.  A corresponding partner
signal was injected into the second scan of this source.  The starting
frequency of the partner signal was obtained by linearly extrapolating
the frequency of the signal in the first scan, i.e., by adding the
product of the artificial drift rate and the known time difference
between the two scans.  The drift rate of the partner signal was set
equal to the that of the original signal plus an increment randomly
chosen from the set $\{- \Delta \dot f, 0, \Delta \dot f\}$.

\begin{figure}[h!]
  \plotone{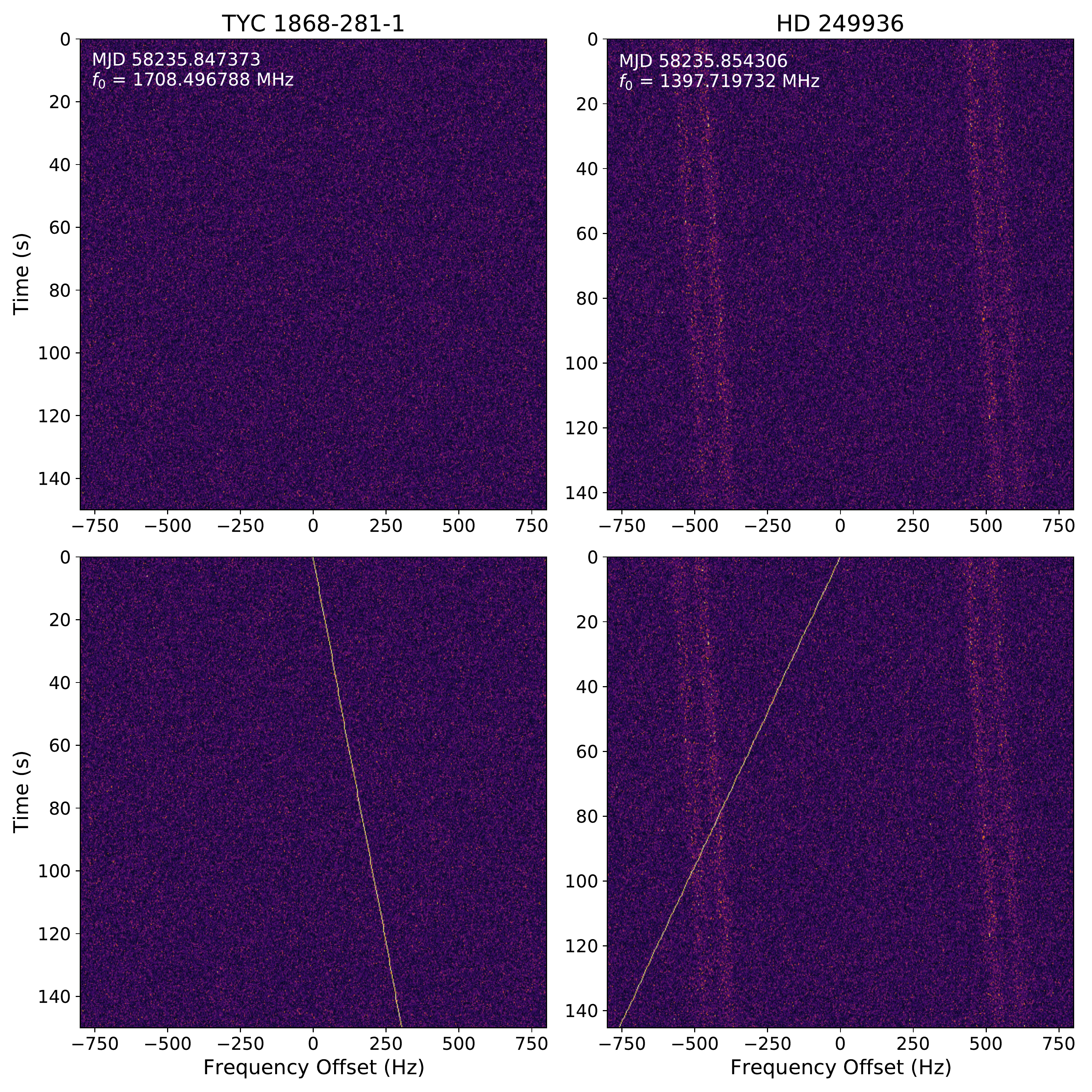}
  \caption{(Top) Time-Frequency diagram before signal injection.
  (Bottom) Time-Frequency diagram after signal injection.  The
  injected signal S/N was increased twentyfold to facilitate
  visualization.  The bottom left panel shows a signal that was
  successfully recovered by our data processing pipeline.  The
  injected signal in the bottom-right panel crosses a stronger RFI
  signal and was missed by our detection algorithm.
  \label{fig:Injected_signal_example}}
\end{figure}

We injected half of the signals at our detection threshold
($10\sigma$) to test the limits of our pipeline's detection
capabilities.  The remaining signals were injected at an S/N of
$20$ to test our sensitivity to stronger signals.  A total of
\injectedcount signals were injected into the April 27, 2018 data.  A
full list of the injected signal properties is available
as supplemental online material.
Two examples of
injected signals are shown in Figure
\ref{fig:Injected_signal_example}\replaced{.}{ and listed in Table~\ref{tab-inj}}.

\begin{table}[h]
\begin{center}
  \begin{tabular}{rrrrrrr}
    NAME & SCAN & FREQ (Hz) & DFDT (Hz/s) & SNR & DETECTED & DOO\_CORRECT\\
    \hline
  TYC 1868-281-1 & 1 & 1708496788.1 & 2.029626 & 10 & Y & Y\\
  HD 249936 & 1 & 1397719731.9 & -5.221517 & 20 & N & N/A\\
  \end{tabular}
  \caption{Properties of artificial signals used for the signal
    injection and recovery analysis.  Columns show source name, scan
    number, frequency of injection at the start of the scan, frequency
    drift rate, signal-to-noise ratio, boolean indicating whether the
    signal was recovered by the pipeline, boolean indicating whether the
    direction-of-origin filter made the correct assignment.  (This
    table is available in its entirety in a machine-readable form in
    the online journal. A portion is shown here for guidance regarding
    its form and content.)}
\end{center}
  \label{tab-inj}
\end{table}

\subsection{Recovery and Classification of Injected Signals} \label{subsec:injection_results}

\begin{figure}[h!]
  \plotone{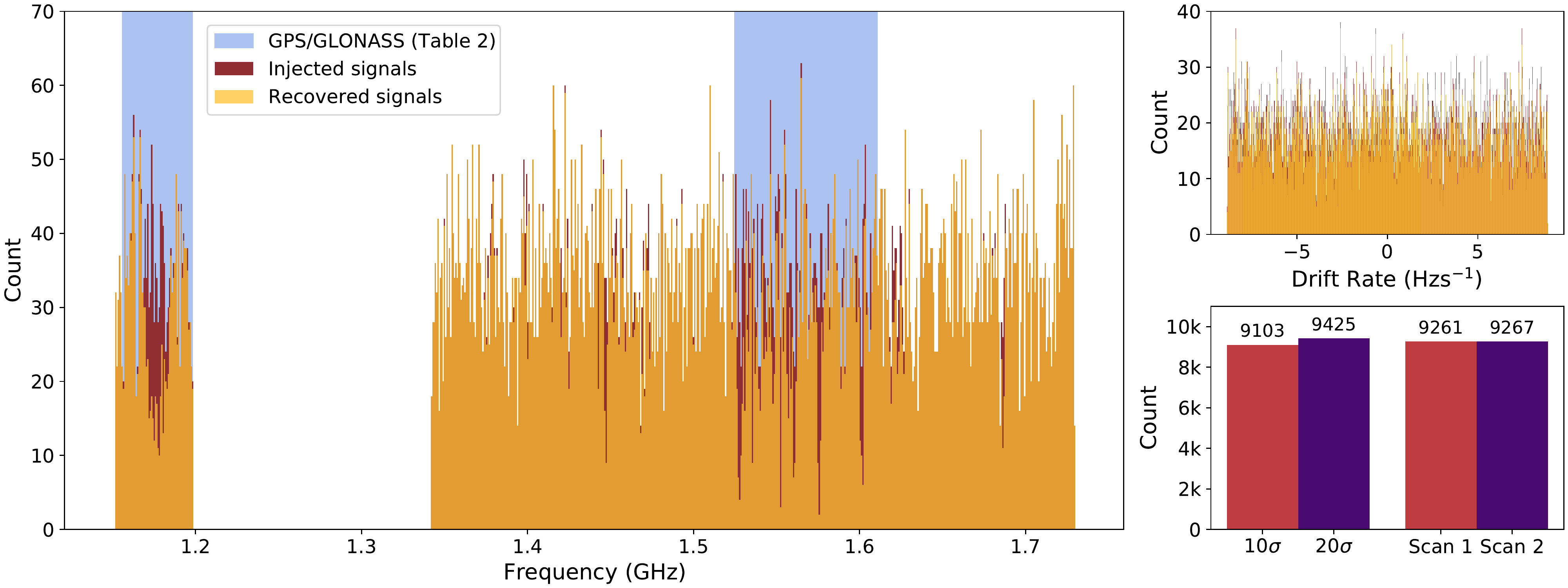}
\caption{(Left) Frequency distribution of injected and recovered
  signals.  The number of signals recovered within known RFI regions
  (such as GPS or GLONASS) is substantially lower than in other
  regions.  (Top right) Drift rate distribution of injected and
  recovered signals.  (Bottom right) Signals recovery counts as a
  function of S/N and scan number.  We observe no significant difference in
  the recovery rate as a function of drift rate or scan number, but we
  do notice a $\sim$3\% increase in the recovery rate of signals with
  larger S/N.
\label{fig:RecoveryStats}}
\end{figure}

After injecting the signals into the dynamic spectra, we applied our
candidate detection procedure (Section
\ref{subsec:candidate_signal_detection}) and stored the output in a
SQL database.
Signals were considered properly recovered if their properties matched those of
the injected signals within $\pm2$ Hz in frequency, $\pm\Delta \dot f$
in drift rate, and $\pm0.1$ in S/N.
We found that our procedure recovered
\recoveredcount(\recoveredpercent) of the injected signals.  Outside
of the
regions with dense RFI described in Table \ref{tab:known_RFI}, our
pipeline performs better, with a recovery rate of
\recoverednoRFIpercent.  We observe no significant difference in the
recovery rate as a function of drift rate or scan number (Figure
\ref{fig:RecoveryStats}), but we do notice a $\sim$3\% increase in the
recovery rate for signals with larger S/N.

We found that most of the signals missed by our pipeline were injected
in regions of known RFI (Figure \ref{fig:RecoveryStats}).  This
pattern is a consequence of two known limitations of our candidate
detection procedure.  First, our algorithms only detect the signal
with highest S/N when two signals intersect in time-frequency space
\citep{Pinchuk2019}.  Second, signals with a low prominence
superimposed on an elevated noise baseline are discarded (Section
\ref{subsec:candidate_signal_detection}).  High-density RFI regions
such as the ones listed in Table \ref{tab:known_RFI} are conducive to
both of these conditions, thereby reducing the recovery rate.
A cursory analysis suggests that $\sim$70--80\% of the non-detections
are due to the intersecting condition.

In order to quantify the performance of our Doppler and
direction-of-origin filters (Section \ref{subsec:SQL_filters}), we
applied our filters to the entire set of detected signals, including
the detections resulting from injected signals.  To distinguish the
performance of these filters from that of our detection algorithm, we
removed \recoverednopartnercount of the injected signals that were
detected in only one scan of a source.  Furthermore, we removed
\recoverednopartnerdfdtnullcount signals that were injected with a
Doppler drift rate of 0 (i.e., stationary with respect to the
observer).  Of the remaining \recoveredshouldbeFANDYcount injected
signals, \recoveredareFANDYcount (\recoveredareFANDYpercent) were
flagged as promising technosignature candidates by our Doppler and
direction-of-origin filters.

\subsection{Performance of Data Processing Pipeline} \label{subsec:injection_discussion}

The preliminary injection and recovery analysis described in this
section identified some important limitations of our radio
technosignature detection pipeline.
Our detection algorithm, which is an improvement over those of
\citet{Margot2018} and \citet{Pinchuk2019} (Figure
\ref{fig:signal_counts_detection}) and outperforms those of
\citet{Enriquez2017} and \citet{Price2020} (Section
\ref{subsec:cand_detection_comparison}), experiences degraded
performance in
regions with dense RFI.
In these regions, it is more
likely for a technosignature candidate to intersect a strong RFI
signal (Figure \ref{fig:Injected_signal_example}), thereby escaping
detection by our pipeline.  This limitation could be overcome by using
the recorded drift rates and starting frequencies of two signals
within a scan to determine whether the signals are predicted to
intersect each other in the other scan of the source.  If an intersection
condition were detected, the known signal could be blanked or replaced
with noise and a new detection procedure could be run to identify
previously undetected signals.
In the presence of strong RFI, a fraction of the injected signals
escape detection because their prominence is below our detection
threshold (i.e., prominence $<$ $f \times $ integrated power, with
$f$=75\%).
In some situations, a valid technosignature could also be removed if
it were detected in a frequency region corresponding to a broadband
signal.  It may be possible to overcome this limitation in the future
by including a comparison of the properties of the narrowband signal
(e.g., drift rate, modulation, etc.) to those of the underlying
broadband signal.

Our improved Doppler and direction-of-origin filters performed
exceptionally well, only mislabeling 49 of the
\recoveredshouldbeFANDYcount injected signals.  The signals that were
incorrectly flagged were paired with an RFI signal of similar drift
rate in a scan of a different source.  This issue can be mitigated by
expanding the signal matching criteria to include signal properties
other than starting frequency and drift rate, such as bandwidth or
gain ratio.

The results presented in this section provide important insights into
the detection capabilities of our current data processing pipeline.
In particular, they demonstrate that our pipeline still misses some of
the narrowband signals that it is designed to detect.  These results
are also useful to identify specific areas in need of improvement.

\subsection{Limitations of Current Signal Injection and Recovery Analysis} \label{subsec:injection_limitations}
The analysis presented in this section is preliminary because it
injects signals into the dynamic spectra and not into the raw data.
Therefore, the current implementation does not consider certain data
processing steps such as correcting for the bandpass channel response
(Section \ref{subsec:preproc}), calculating the noise statistics and
normalizing the power spectra to zero mean and unit variance, or
applying
the incoherent dechirping procedure (Section
\ref{subsec:treealg}).

In future work, we will implement the capability to inject signals in
the raw data.  This improved implementation will allow us to quantify
the detection performance of the entire pipeline.  We anticipate that
it will also be helpful in revealing additional areas for improvement.

\section{Results}\label{sec:results}

We applied the methods described in Section \ref{sec:analysis} to the
data described in Section \ref{subsec:sources}.  We detected a total
of \totcount candidate signals over both 2018 and 2019 observation epochs.
We used the total integration time of 2.75~hr and
\replaced{useful bandwidth of $\Delta f_{\rm tot} = 309.3$ MHz}{processed bandwidth of 438.8 MHz} to compute a signal detection count
per unit bandwidth per unit integration time.  In BL parlance, our
detections are referred to as ``hits'' \citep{Enriquez2017,Price2020}
and the hit rate density of this search is
\replaced{3.1}{2.2} $\times 10^{-2}$ hits per hour per hertz.  In comparison, the
L-band component of \citet{Price2020}'s search with the same telescope
and S/N threshold resulted in 37.14 million hits in 506.5 hr over a
useful bandwidth of 660 MHz, or a hit rate density of 1.1 $\times
10^{-4}$ hits per hour per hertz, \replaced{almost 300}{200} times smaller than ours.  We
discuss possible reasons for this large differential in Section
\ref{subsec:cand_detection_comparison}.

A complete table of signal properties of the detected candidates is
available online\footnote{\url{https://doi.org/10.5068/D1937J}}.
Our
Doppler and direction-of-origin filters flagged \filterremovedcount
(\filterremovedpercent) signals as
anthropogenic RFI.  A majority of the remaining \postfiltercount
signals were detected within operating regions of known interferers
(Table \ref{tab:known_RFI}).  Candidate signals remaining within these
frequency regions were attributed to RFI and removed from
consideration.

\begin{figure}[h!]
  \plotone{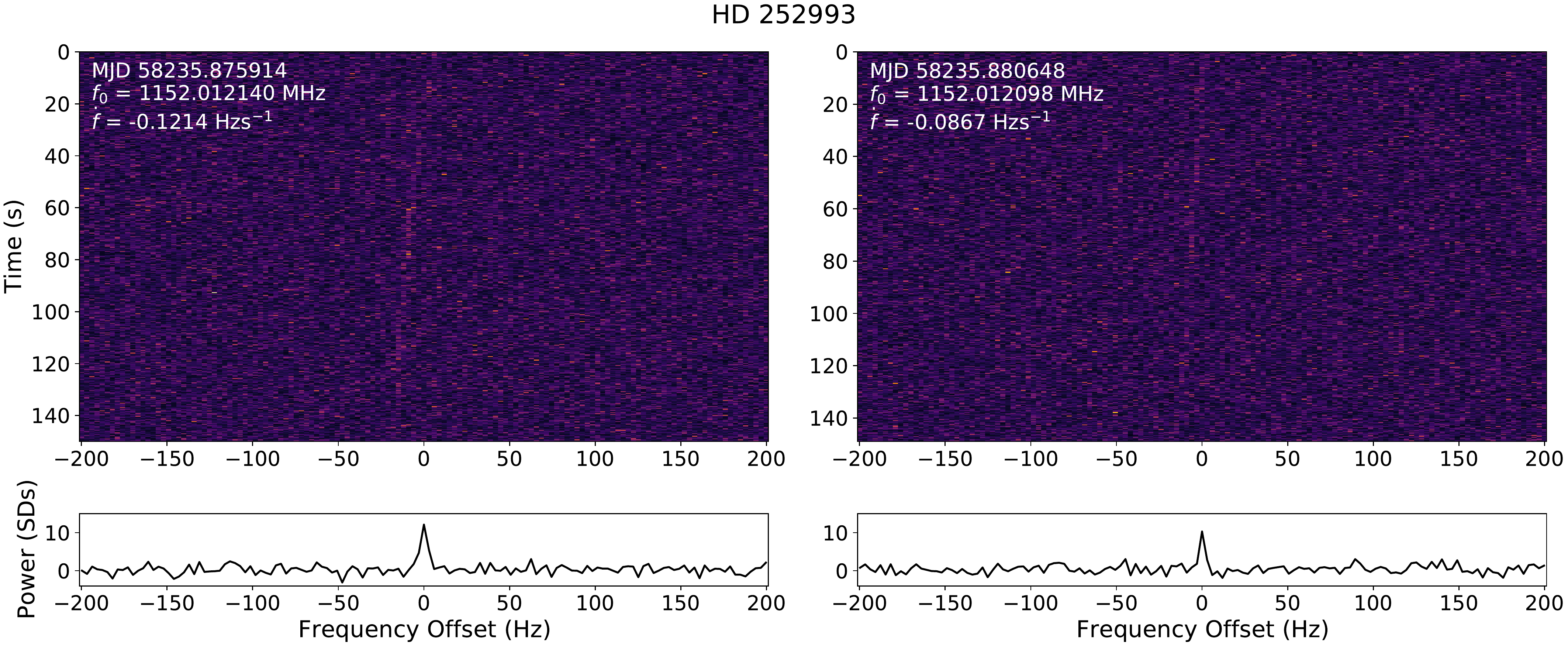}
  \caption{%
    Dynamic spectra (top) and integrated power spectrum
  (bottom) of a final candidate signal that appears in scan 1 (left)
  and scan 2 (right) of HD 252993.  Although this signal exhibits many
  of the desirable properties of a technosignature (e.g., narrowband,
  non-zero Doppler drift rate, persistence), it was ultimately
  rejected because it was visually confirmed to appear in multiple
  directions on the sky.
  \label{fig:example_signal}}
\end{figure}

The remaining \finalcount signals were deemed most promising
technosignature candidates.  Visual inspection of all of these
candidates revealed that they are attributable to RFI.  Figure
\ref{fig:example_signal} shows an example of a promising signal that
was ultimately attributed to RFI.

The vast majority
of the most promising candidates were eliminated because they were
detected in multiple directions on the sky.  These signals escaped
automatic RFI classification by our filters for one or more of the
following reasons, which are generally similar to the ``Categories''
described by \citet[][Section 4]{Pinchuk2019}:
\begin{enumerate}
  \item The S/N values of corresponding signals in scans of other sources were below the detection threshold of 10.
    This difficulty could perhaps be circumvented in the future by conducting an additional search for lower S/N signals
    at nearby frequencies.
  \item The drift rate of the signal differed from those of
    corresponding signals in scans of other sources by more than our
    allowed tolerance ($\pm\Delta \dot{f}$ = $\pm0.0173$ \Hzsns).
  \item The signal was not detected in scans of other sources because
    it intersected another signal of a higher S/N.
  \item The signal bandwidth exceeded 10 Hz, making it difficult to
    accurately determine a drift rate and therefore link the signal with
    corresponding signals in scans of other sources.
\end{enumerate}
All of these difficulties could likely be overcome by a
direction-of-origin filter that examines the time-frequency data
directly instead of relying on estimated signal properties such as
starting frequency and drift rate.
We are in the process of implementing machine learning tools for this purpose.

Because automatic classification and visual inspection attributed all
of our candidate signals to RFI, we did not detect a technosignature
in this sample.
We are preserving the raw data in order to enable reprocessing of the
data with improved algorithms in the future, including searches for
additional types of technosignatures.

\section{Discussion}\label{sec:discussion}

\subsection{Dechirping Efficiency} \label{subsec:drift_rates_comparison}

Over sufficiently short ($\sim$5 min) scan durations, monochromatic signals emitted
on extraterrestrial platforms are well approximated by linear chirp
waveforms ($f(t) = f(t_0) + \dot{f} (t - t_0)$).
Most radio technosignature detection algorithms rely on incoherent
dechirping, i.e., incoherent sums of power spectra, to integrate the
signal power over the scan duration (Section \ref{subsec:treealg}).
In the context of incoherent sums, the magnitude of the
maximum drift rate that can be considered without loss in sensitivity
is given by
\begin{equation} \label{eq:max_drift_rate}
  \dot f_{\rm max} = \frac{\Delta f}{\Delta T},
\end{equation}
where $\Delta f$ is the adopted spectral resolution and $\Delta T$ is
the accumulation time corresponding to one row in the dynamic spectra.  If the drift rate
of a signal exceeds this maximum drift rate ($\dot f > \dot f_{\rm
  max}$), the signal frequency drift exceeds ${\Delta f}$ during 
  ${\Delta T}$, and power is smeared over multiple
frequency channels,
resulting in reduced sensitivity.
\enlargethispage{1cm}
In this work ($\Delta f = 2.98$~Hz; $\Delta T = 1/{\Delta f} =
0.34$~s), the maximum sensitivity can be obtained up to frequency
drift rates of $\dot f_{\rm max, UCLA} = 8.88$~\Hzsns.
BL investigators \citet{Enriquez2017} and \citet{Price2020} used
$\Delta f = 2.79$~Hz and $\Delta T = 51/{\Delta f} = 18.25$~s, which
yields $\dot f_{\rm max, BL} = 0.15$~\Hzsns.  However, these authors
conducted searches for signals with drift rates larger than
0.15~\Hzsns, resulting in reduced sensitivity for $>$90\% of the drift
rates that they considered.
For instance, at the largest drift rate considered by
\citet{Price2020}, the frequency drifts by 4 \Hzs $\times$ 18.25 s =
73 Hz (26 channels) during $\Delta T$, and only $\sim$4\% of the
signal power is recovered in each frequency channel.
We express this loss of signal power with a detection
efficiency in the range 0--100\% and refer to it as a {\em dechirping
  efficiency}.

\begin{figure}[ht!]
\begin{tabular}{cc}
  \includegraphics[width=4in]{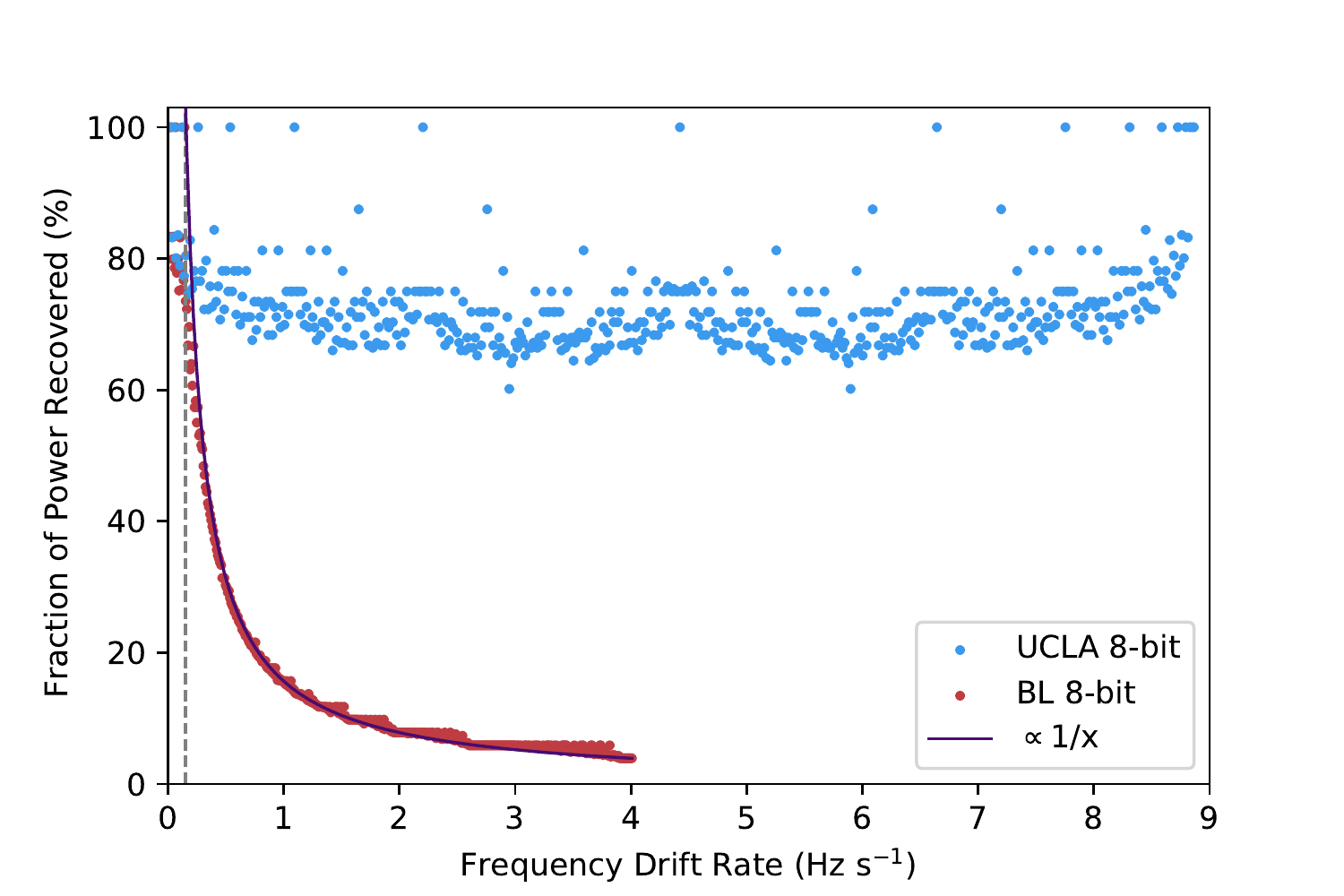} &
  \includegraphics[width=2.5in]{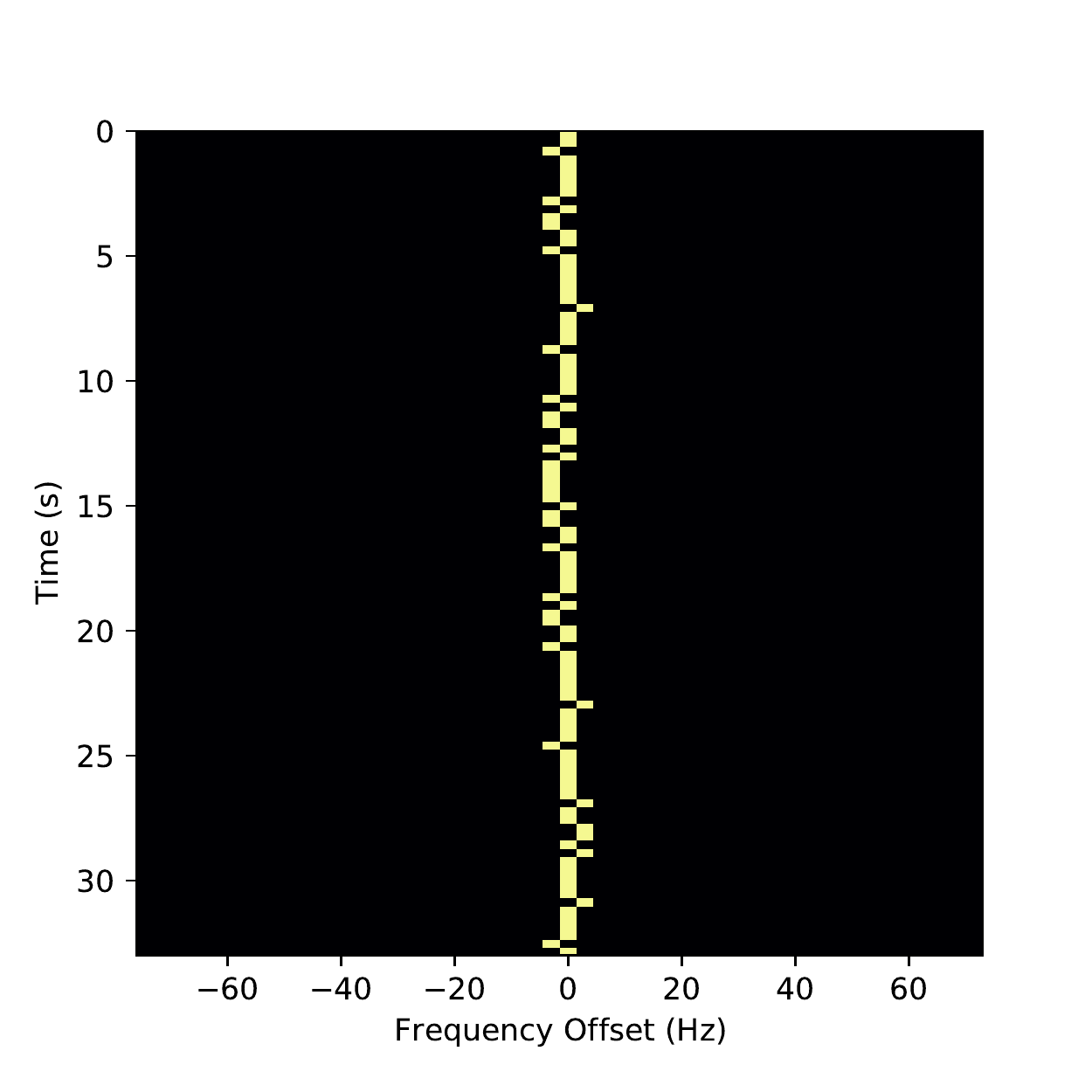}\\
\end{tabular}
\caption{ (Left)
  Dechirping efficiencies
  of the UCLA (blue) and BL (red) data
  processing pipelines as a function of Doppler frequency drift rate
  at the nominal frequency resolutions of $\sim$3 Hz.
  Our choices of data-taking and data-processing parameters result in
  a fairly uniform efficiency (72.4\% $\pm$ 6.8\%) across the full
  range of drift rates considered, with values below 100\% due to
  imperfections of the tree algorithm (see text).  The BL choices
  result in considerably reduced detection efficiency beyond $\dot
  f_{\rm max,BL} = 0.15$~\Hzsns (dashed vertical line), with values as
  low as 4\% due to smearing of the signal power across multiple
  frequency bins.  The performance at frequencies beyond $\dot f_{\rm
    max}$ is well approximated by a $1/x$ function (purple line),
  consistent with the inverse bandwidth dependence of the amplitude of
  a linear chirp power spectrum.  (Right) Dynamic spectrum of a linear
  chirp waveform dechirped imperfectly by the tree algorithm.  In
  this worst-case scenario
  for $\dot f \leq \dot f_{\rm max}$, only 60\% of the spectra are shifted by
  the correct amounts and only 60\% of the power is recovered in the
  appropriate frequency channel.  
  Only the first 100 rows ($\sim$30 s) are shown.
  \label{fig:dfdt_comparison}}
\end{figure}

To confirm the performance of the data processing pipelines, we
conducted numerical experiments\footnote{Our software is available at \href{https://github.com/UCLA-SETI-Group/dechirping_efficiency}{https://github.com/UCLA-SETI-Group/dechirping\_efficiency.}}
with both our algorithms and BL's {\em
  turbo}SETI package \citep{Enriquez2017}.  For the purpose of these
simulations, we created noise-free, constant-power dynamic spectra of
linear chirp waveforms with the frequency and time resolutions
appropriate for the UCLA and BL searches.  By considering only
integral pixel locations, we simulated
frequency drift rates that are exact
multiples of the elemental drift rates considered by our respective
tree algorithms (0.0173~\Hzs for the UCLA searches, 0.0096~\Hzs for
the BL searches).
We ran the respective tree algorithms on the simulated spectra and
recorded the power recovered at each drift rate as a function of total
signal power (Figure \ref{fig:dfdt_comparison}, Left).  The
experiments show that,
at the nominal frequency resolutions of $\sim$3 Hz,
dechirping efficiencies of 100\% are possible in our and other
searches with $\dot f \leq \dot f_{\rm max}$, whereas dechirping
efficiencies rapidly degrade to values as low as 4\% in the BL
searches with $\dot f > \dot f_{\rm max,BL}$

In this experiment, a perfect algorithm would recover 100\% of the
signal power, as long as $\dot f \leq \dot f_{\rm max}$.  The tree
algorithm (Section \ref{subsec:treealg}) is not perfect in that it
reuses pre-computed sums to achieve $N \log N$ computational cost.  As
a result, the tree algorithm shifts every spectrum
by an amount that is not always optimal.  In other words, it is unable to perfectly
dechirp most linear chirp waveforms.
In our simulations of the UCLA pipeline, we do observe 100\% of the
power recovered for several drift rates (Figure
\ref{fig:dfdt_comparison}, Left).  On average, the pipeline recovers
72.4\% $\pm$ 6.8\% of the signal power.  In the worst-case scenario,
the fraction of power recovered is 60\%.  The tree algorithm's
dechirped waveform of this worst-case scenario reveals that 60\% of
the frequency bins are shifted to the correct locations and 40\% are
shifted to incorrect locations (Figure \ref{fig:dfdt_comparison},
Right).
We quantified the dechirping efficiencies associated with the use of
the tree algorithm for a variety of array dimensions (Table~\ref{tab:dechirp}).

We computed a rough estimate of the mean dechirping efficiency in the
search of \citet{Price2020} for
the nominal frequency resolution of $\sim$3 Hz and 
a uniform distribution of candidate
signals as a function of drift rate.  We assumed a generous 100\%
efficiency between 0 and 0.15 \Hzs and the 1/x trend observed in
Figure \ref{fig:dfdt_comparison} between 0.15 \Hzs and 4 \Hzsns.  We
found a mean efficiency of 16.5\%.  A weighted mean of the efficiency
based on the exact distribution of signals as a function of drift rate
would provide a more accurate and likely larger value.

\begin{deluxetable}{rrrrrr}[h!]
  \tablehead{\colhead{Rows} & \colhead{Min} & \colhead{Max} & \colhead{Mean} & \colhead{Median} & \colhead{STD}}
  \tablehead{\colhead{Rows} & \colhead{Min (\%)} & \colhead{Max (\%)} & \colhead{Mean (\%)} & \colhead{Median (\%)} & \colhead{STD (\%)}}
\startdata
     4 & 100.00  & 100.00  & 100.00  & 100.00  &   0.00  \\
     8 &  75.00  & 100.00  &  93.75  & 100.00  &  11.57  \\
    16 &  75.00  & 100.00  &  90.62  &  93.75  &  10.70  \\
    32 &  68.75  & 100.00  &  85.16  &  81.25  &  11.20  \\
    64 &  68.75  & 100.00  &  81.64  &  78.12  &   9.83  \\
   128 &  64.06  & 100.00  &  77.93  &  75.00  &   8.88  \\
   256 &  64.06  & 100.00  &  75.17  &  73.44  &   7.68  \\
   512 &  60.16  & 100.00  &  72.42  &  71.09  &   6.84  \\
  1024 &  60.16  & 100.00  &  70.08  &  69.14  &   6.10  \\
  2048 &  56.84  & 100.00  &  67.92  &  66.60  &   5.53  \\
  4096 &  56.84  & 100.00  &  66.01  &  64.94  &   5.06  \\
  \enddata
  \caption{Dechirping efficiencies resulting from incoherent
    dechirping of power spectra with a computationally advantageous
    but approximate tree algorithm (Section \ref{subsec:treealg}).}
  \label{tab:dechirp}
\end{deluxetable}

We describe two alternate, partial solutions to the loss of
sensitivity sustained during incoherent dechirping.  The first is to
reduce the frequency resolution of the dynamic spectra, thereby
increasing the range of drift rates that can be explored without
spreading power across multiple channels \citep[e.g.,][]{Siemion2013}.
However, this solution still results in loss of sensitivity.  For
narrowband signals, each doubling of the frequency resolution results
in a $\sqrt{2}$ decrease in sensitivity.  To reach the maximum drift
rates of $\pm 4$~\Hzsns considered by \citet{Price2020}, one would
have to apply 4--5 doublings, resulting in frequency resolutions of
45--90~Hz and sensitivity to narrowband signals of 18--25\% of the
nominal value.  Another, related approach, would be to use a
drift-rate-dependent boxcar average of the integrated spectra to
recover the power that has been spread over multiple channels, e.g.,
by averaging 26 channels at the maximum drift rates of $\pm 4$~\Hzsns
considered by \citet{Price2020}.  Doing so would degrade the frequency
resolution to values up to 73~Hz and the sensitivity to narrowband
signals to 20\% of the nominal value.

\subsection{Extreme Drift Rates} \label{subsec:extreme}

In a recent study\footnote{\citet{Sheikh2019} incorrectly delineated
  the search parameters of \citet{Enriquez2017} in their Figure 1.
  The maximum frequency excursion considered in that search is 600 Hz,
  not 12,000 Hz.}  of the expected drift rates of a large class of
bodies, including exoplanets with highly eccentric orbits and small
semimajor axes, \citet{Sheikh2019} recommended searching drift rates
as large as $\dot f / f_{obs} = 200$~nHz At the center frequency of
our observations (1.5 GHz), this corresponds to a drift rate of
300~\Hzsns.  %
Our data archival policy (Section \ref{subsec:data_discussion}) would
enable reprocessing of the data with parameters that are more
conducive to large drift rates.  For example, we could reprocess our
data with Fourier transforms of length $2^{17}$.  This choice would
increase our frequency resolution 8-fold to 24 Hz and would allow us
to search for drift rates up to $\sim 570$~\Hzs without incurring any
sensitivity loss due to signal smearing over multiple frequency
channels.
In contrast, BL archive products include dynamic spectra but do not
include most of the raw voltage data
\citep{lebo19,Enriquez2017,Price2020}, making it impractical to
conduct a search with archival products at drift rates larger than
$\sim$1~\Hzs with adequate sensitivity (Figure \ref{fig:dfdt_comparison}).

\subsection{Data Requantization and Preservation} \label{subsec:data_discussion}

\begin{figure}[ht!]
\plotone{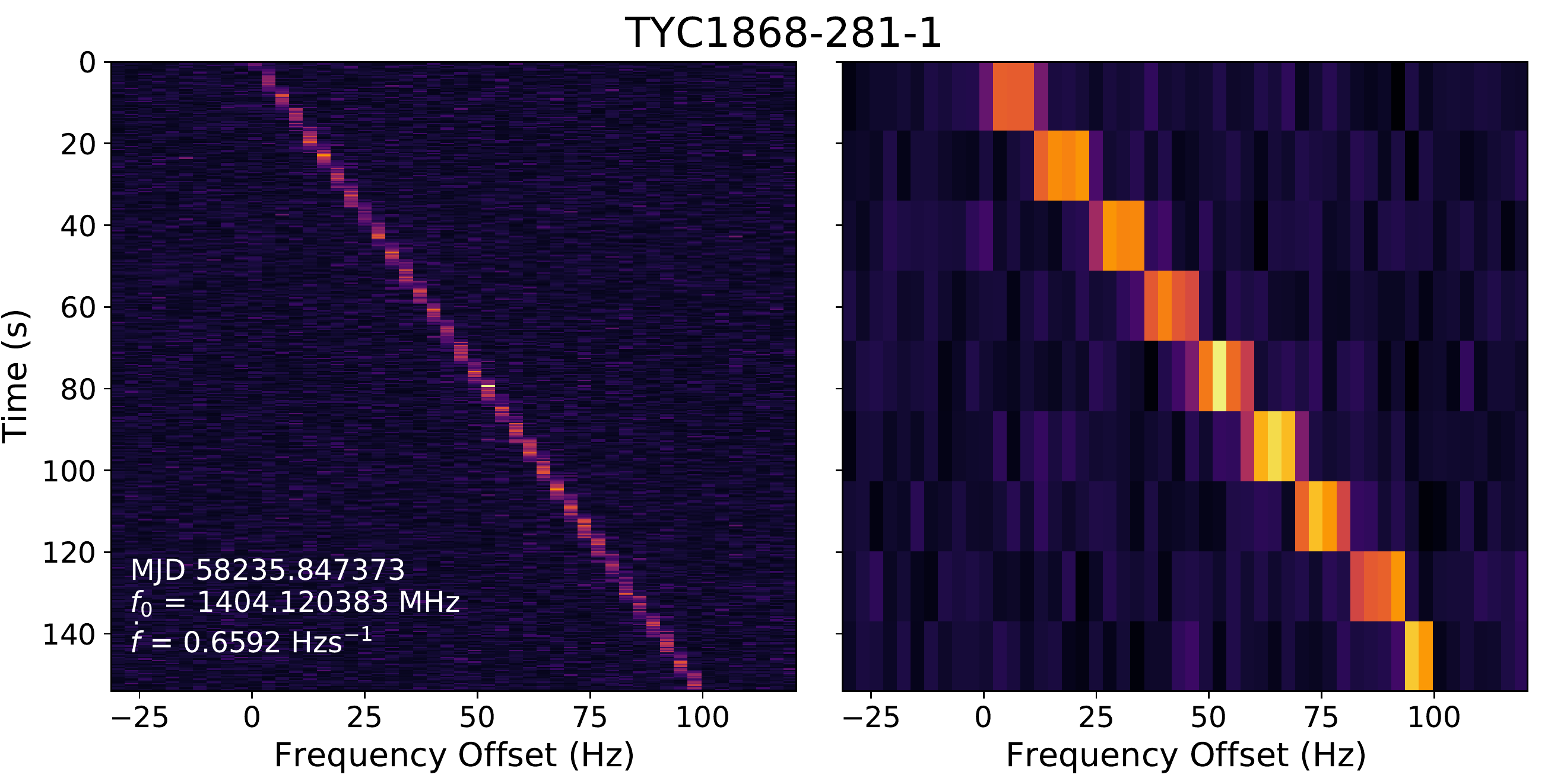}
\caption{
  Representative dynamic spectra of a signal shown with
  the nominal time resolution of $\sim$1/3 Hz = 0.33 s (left)
  and with the degraded time resolution 
  resulting from time-averaging
  51 consecutive spectra
  (right).
   \label{fig:compare_time_res}}       
\end{figure}

Our choice of data recording parameters is largely driven by our
dedication to preserve the raw voltage data recorded during our
observations.
We prefer to archive the raw data as opposed to derived data products
such as dynamic power spectra, for four reasons.  First, the raw 2-bit
data require less storage space than the 32-bit dynamic spectra.  
Second, the dynamic power spectra can be easily regenerated from the
raw data, but the reverse is not true, because phase information is
lost in the process of computing power spectra.  Third, there are
large penalties associated with preserving incoherent averages of
individual power spectra.  \citet{Enriquez2017} and \citet{Price2020}
average 51 consecutive spectra to keep the archival volume manageable,
which degrades the sensitivity of the search by factors of up to
$\sim$25 (Section \ref{subsec:drift_rates_comparison}) and the time
resolution by a factor of 51 (Figure \ref{fig:compare_time_res}).
As a result, the BL dynamic spectra would not be useful in confirming
or interpreting a signal with 1 Hz modulation, for instance.  Fourth,
the only way to preserve the ability to conduct novel or improved data
analysis with maximum sensitivity and resolution
is to preserve the raw data.
However, there are penalties associated with storing raw data in 2-bit
format as opposed to 8-bit format \citep[e.g.,][]{Price2020}.

For this work and previous analyses \citep{Margot2018,Pinchuk2019}, we
selected a data-taking mode that yields 2-bit raw voltage data after
requantization with an optimal four-level sampler \citep{Kogan1998}.
The quantization efficiency, which is the ratio of signal power that
is observed with the optimal four-level sampler to the power that
would be obtained with no quantization loss, is 0.8825.
\citet{Price2020} noted that a consequence of this requantization is
that the S/N threshold used in this work (10) would need to be
lowered by approximately 12\% to detect the same number of candidate
signals as 8-bit quantized data.  While we agree with this statement,
the S/N threshold of radio technosignature searches is somewhat
arbitrary, and our choice compares favorably to that of other surveys
(Table \ref{tab:SNR_comparison}).  Should the need ever arises to
detect weaker signals, we would simply re-analyze our data with a
lower S/N threshold.
In addition, the sensitivity enabled by our decision to minimize the
accumulation time when computing dynamic spectra 
(Section \ref{subsec:drift_rates_comparison}) offsets the losses due
to quantization efficiency compared to pipelines with longer
accumulation times.  Specifically, if we apply the 0.8825 quantization
efficiency to the results illustrated in
Figure~\ref{fig:dfdt_comparison},
we find that our overall sensitivity surpasses BL's sensitivity for
any drift rate larger than
0.153~\Hzs
and surpasses it by a factor of at least 5 for any drift rate
larger than
1.11~\Hzs.

\begin{deluxetable}{lc}[ht!]
  \caption{S/N thresholds used in recent searches for radio technosignatures.
\label{tab:SNR_comparison}}
\tablehead{
\colhead{Reference} & \colhead{S/N} 
}
\startdata
\citet{Gray2017}              & 7     \\ %
\citet{Harp2016}              & 9/6.5 \\
UCLA SETI searches            & 10    \\ 
\citet{Price2020}             & 10    \\
\citet{Enriquez2017}          & 25    \\
\citet{Siemion2013}           & 25    \\
\enddata
\end{deluxetable}

\subsection{Candidate Signal Detection Count} \label{subsec:cand_detection_comparison}

Our results indicate a hit rate density of \replaced{3.1}{2.2} $\times 10^{-2}$ hits
per hour per hertz whereas \citet{Price2020} obtained a considerably
lower value of 1.1 $\times 10^{-4}$ hits per hour per hertz with the
same telescope and S/N threshold (Section \ref{sec:results}).  We
investigate possible causes for this factor of $\sim$\replaced{300}{200} difference.
First, our observing cadence involves two scans of 150 s each per
source, whereas \citet{Price2020} used three scans of 300 s each per
source.  The difference in integration time could perhaps be invoked
to explain a factor of up to 3 difference in hit rate density,
although a larger number of signals ought to be detectable with BL's
longer scan durations.  Second, our \replaced{usable}{processed} frequency range extends
over \replaced{309.3}{438.8} MHz, whereas \citet{Price2020} used
a superset of that range extending over 660 MHz.  A non-uniform
distribution of dense RFI across the spectrum could perhaps be invoked
to explain a factor of up to $\sim$2 difference in hit rate density.
Third, we examine a range of drift rates that is twice as large as the
range used by \citet{Price2020}, which may explain a factor of $\sim$2
difference in hit rate density if the distribution of hits as a
function of drift rate is roughly uniform.
These
small factors cannot explain the two orders of magnitude difference in
hit rate density, which must be related to more fundamental effects.
We surmise that the two most important factors are the difference in
the effective sensitivities of our searches due to different
dechirping efficiencies (Section \ref{subsec:drift_rates_comparison})
and the algorithmic difference in the identification of candidate
signals or hits.

As detailed by \citet{Pinchuk2019}, the candidate signal detection
procedures used in several previous radio technosignature searches
\citep[e.g.,][]{Siemion2013, Enriquez2017, Margot2018, Price2020}
unnecessarily remove kHz-wide regions of frequency space around every
signal detection.
This practice complicates
attempts to place upper
limits on the existence of technosignatures, because the algorithms
discard many signals that are legitimate technosignature candidates.
In addition, this practice leads to slight overestimates of search
metrics, such as the Drake Figure of Merit (DFM) \citep{Pinchuk2019}.
Here, we quantify the number of signals that
are unnecessarily discarded by algorithms that remove $\sim$kHz wide
frequency regions around every detection.

To perform this comparison, we used the database of signals detected
during the 
April 27, 2018 observations 
and we replicated the procedure described by
\citet{Enriquez2017} and \citet{Price2020}.  The ``blanking''
procedure used by \citet{Price2020} specifies ``Only the signal with
the highest S/N within a window ... $\pm$ 600~Hz is recorded as a
hit.''  To replicate this step, we sorted the signals detected in each
scan in decreasing order of S/N and iterated over the sorted lists.
At every iteration, we kept the signal with the largest remaining S/N
value and eliminated all other signals
within $\pm$ 600~Hz.
The next step described by \citet{Price2020} combines hits that fall
within a certain frequency range into groups as long as the signal is
detected in every scan of the source.  We replicated this step by
grouping signals that were present in both scans of each source
according to \citet{Price2020}'s prescription for frequency range.
The third step of the procedure described by \citet{Price2020} reads:
``Additionally, any set of hits for which there is at least one hit in
the OFF observations within $\pm$ 600~Hz of the hit frequency from the
first ON observation would be discarded.''  This elimination seems
wasteful because the presence of OFF-scan signals with
drift rates that are unrelated to the ON-scan drift rate
results in elimination.  To replicate this step, 
we removed all groups of signals for which one or both of the two OFF-scans 
contained an unrelated signal
within $\pm$ 600~Hz of the detection in the first ON-scan.
To determine whether signals were unrelated, we placed the following
condition on drift rate:
\begin{equation}\label{eq:dfdt_tolerance}
|\dot f - \dot f_0| > 2 \Delta \dot{f},
\end{equation}
where $\dot f$ is the drift rate of the OFF-scan signal, $\dot f_0$ is the drift rate of the ON-scan signal, and $\Delta \dot{f} = 0.0173$ \Hzs is the drift rate resolution.
Our direction-of-origin filters (Section \ref{subsec:SQL_filters})
also remove signals if they are found in multiple directions on the
sky, but only in conjunction with careful analysis of the drift rates
of both signals.  Specifically, our filters only remove the two
signals if their drift rates are within a tolerance of $2 \Delta
\dot{f}$.  For the purpose of this blanking analysis, we kept signals
that satisfy this criterion because both pipelines remove them during
subsequent filtering.
We found that our pipeline detected
10,113,551
signals whereas our
pipeline with a blanking algorithm modeled after the descriptions
given by \citet{Enriquez2017} and \citet{Price2020} detected only
1,054,144
signals.  In other words, our pipeline detects 
$\sim$10 
times
as many signals as the BL-like pipeline over the same frequency range,
with a corresponding increase in hit rate density.

To summarize, we found that our algorithmic approach to signal
identification explains the largest fraction of the factor of
$\sim$\replaced{300}{200} difference in hit rate density between our and the BL
searches, likely followed by our better overall sensitivity for 
$>$90\% of the frequency drift rates examined by the BL pipeline
(Section \ref{subsec:drift_rates_comparison}), likely followed by our
shorter integration times and consideration of a wider range of drift rates.
It is also possible that the limited dynamic range of our 2-bit
voltage data makes our search susceptible to spurious detections at
harmonics of strong RFI signals (Danny Price, pers.\ comm.).  We are
planning to quantify the importance of this effect in the future.

The differential in hit rate density has
implications for the
validity of existence limit estimates and figure of merit calculations
described by \citet{Enriquez2017} and \citet{Price2020}.

\subsection{Existence Limits} \label{subsec:existence_limits_discussion}

We describe three issues that affect recent claims about the prevalence
of transmitters in the Galaxy \citep{Enriquez2017,Price2020,wlod20}.

First, the range of Doppler drift rates considered in these searches
has been
limited ($\pm 2$~\Hzsns and $\pm 4$~\Hzsns), whereas
transmitters may be located in a variety of settings with
line-of-sight accelerations that would only be detectable at larger
drift rates \citep[e.g.,][]{Sheikh2019}.

Second, these claims invoke transmitters with certain EIRP values that
are calculated on the basis of the nominal sensitivity to non-drifting
signals.  However, the sensitivity to signals drifting in frequency is
demonstrably degraded (Section \ref{subsec:drift_rates_comparison})
with the incoherent dechirping method used in these searches.  The
published EIRP values could be erroneous by factors of up to 25 for
these searches, depending on the drift rate of the putative signal.

Third, our preliminary candidate signal injection and recovery analysis
(Section \ref{sec:injection_analysis}) reinforces the concerns voiced
by \citet{Margot2018} and \citet{Pinchuk2019} about
\citet{Enriquez2017}'s claims.
\citet{Pinchuk2019} argued that an injection and recovery analysis
would demonstrate that a fraction of detectable and legitimate signals
are not identified by existing pipelines, thereby
requiring corrections to 
the claims.
We have shown that our current
pipeline misses $\sim7\%$ of the signals injected into the dynamic
spectra (Section \ref{sec:injection_analysis}).  We surmise that the BL
pipelines used by \citet{Enriquez2017} and \citet{Price2020} miss a
substantially larger fraction of signals that they are meant to detect
because of reduced sensitivity (Section
\ref{subsec:drift_rates_comparison}), time resolution (Section
\ref{subsec:data_discussion}), and detection counts (Section
\ref{subsec:cand_detection_comparison}) compared to our pipeline.

In light of these issues, published claims about the prevalence of transmitters in the Galaxy
\citep[e.g.,][]{Enriquez2017,Price2020,wlod20} almost certainly need
revision.
As mentioned in Section \ref{subsec:injection_limitations}, we are
planning improvements to our signal injection and recovery analysis.
Until this refined analysis is complete, we will not be in a position
to make reliable inferences about the prevalence of radio beacons in
the Galaxy.

\subsection{Drake Figure of Merit} \label{subsec:DFM}

The Drake Figure of Merit \citep{Drake1984} is a metric that can be
used to compare some of the dimensions of the parameter space examined
by different radio technosignature searches.  It is expressed as
\begin{equation} \label{eq:DFM}
  {\rm DFM} = \frac{\Delta f_{\rm tot} \Omega}{F_{\rm det}^{3/2}},
\end{equation}
where $\Delta f_{\rm tot}$ is the total bandwidth observed, $\Omega$
is the total angular sky coverage, and
$F_{\rm det}$ is the minimum detectable flux.  
Assuming unit quantization and dechirping efficiencies, our search
with S/N threshold of 10 is sensitive to sources with flux densities
of 10 Jy and above \citep{Margot2018}.  For consistency with earlier
calculations \citep{Enriquez2017,Price2020}, we have assumed that the
bandwidth of the transmitted signal is 1 Hz, resulting in a minimum
detectable flux
$F_{\rm det} = 10^{-25}$ W\,m$^{-2}$.
The sky coverage of this search is 
$\Omega = 31 \times 0.015 \ {\rm deg}^2 = 0.465$ deg$^2$, i.e., 11~ppm of the entire sky.
The useful bandwidth  is
$\Delta f_{\rm tot} = 309.3$ MHz (Section \ref{subsec:Freq_filters}).
We used these parameters to calculate the DFM associated with this
search and found ${\rm DFM} = 1.11 \times 10^{32}$, where we have used
units of GHz m$^3$ W$^{-3/2}$ for compatibility with \citet{horo93}.
We reanalyzed our 2016 and 2017 data sets (Section
\ref{sec:1617Results}) and recomputed DFM values of
$5.00 \times 10^{31}$
and
$4.71 \times 10^{31}$
for these data sets, respectively, with an
aggregate DFM for our 2016--2019 searches of
$2.08 \times 10^{32}$.
However, we regard these values and all previously published DFM
values with skepticism.

The DFM values published in recent works
do not provide accurate estimates of search volume or performance for
a few reasons.  First, the DFM relies on minimum detectable flux, but
authors have ignored factors that can tremendously affect overall
search sensitivity, such as quantization efficiency ($\sim$88\% for
2-bit sampling) or dechirping efficiency (60--100\% with the tree
algorithm and the parameters of this search, and as low as $\sim$4\%
in the recent BL search described by \citet{Price2020}).
Second, it does not take account of the range of drift rates
considered in a search, which is clearly an important
dimension of the search volume.  Third, it ignores the quality of the
signal detection algorithms, such that two surveys may have the same
DFM even though their data processing pipelines detect substantially 
different numbers of signals
(e.g., the blanking of kHz-wide regions of frequency space described in Section
\ref{subsec:cand_detection_comparison}).  For these reasons, we
believe that DFM values calculated by authors of recent searches,
including our own, are
questionable indicators of actual search volume or performance.
\citet{horo93} expressed additional concerns, stating that 
the DFM ``probably does
justice to none of the searches; it is a measure of the odds of
success, assuming a homogeneous and isotropic distribution of
civilizations transmitting weak signals at random frequencies.''

In Section \ref{subsec:drift_rates_comparison}, we showed that the
dechirping efficiency degrades rapidly for frequency drift rates
larger than $\dot{f}_{\rm max}$ (Figure \ref{fig:dfdt_comparison}).
As a result, the minimum detectable flux for non-drifting signals,
which has been used by \citet{Enriquez2017} and \citet{Price2020} in
their DFM estimates (Equation \ref{eq:DFM}), is not representative of
the minimum detectable flux of signals with $>$90\% of the drift rates
that they considered, which can be up to 25 times larger.
Given the presence of this flux
to the 3/2 power in the denominator of the DFM, we believe that the
DFMs of these searches have been inadvertently but considerably
overestimated.
Other figures of merit, such as \citet{Enriquez2017}'s `Continuous
Waveform Transmitter Rate Figure of Merit' (CWTFM), are also affected by this
problem.

We can use our estimates of the mean dechirping efficiencies to
quantify plausible errors in DFM estimates.  In Section
\ref{subsec:drift_rates_comparison}, we computed a rough estimate of
16.5\% for the mean efficiency of the BL search conducted by
\citet{Price2020}, suggesting that the DFM of their search has been
overestimated by a factor of $\sim$15.  This value may be revised down
once a more accurate estimate of the mean dechirping efficiency
becomes available.  For the UCLA searches conducted between 2016 and
2019, the mean dechirping efficiency is 72.4\% and the quantization
efficiency is 88.25\%, resulting in an overall efficiency of 64\% and
DFM overestimation by a factor of $\sim$2.

\subsection{Other Estimates of Search Volume} \label{subsec:OFM}

The range of drift rates considered in a search
program obviously affects the probability of success of detecting a
technosignature.  For instance, a search restricted to drift rates
smaller than $\dot f_{\rm max, BL} = 0.15$~\Hzs could fail to detect
the signal from an emitter on an Earth-like planet.  The frequency
drift rate dimension of the search volume does not appear to have been
fully appreciated in the literature.  It is distinct from the
``modulation'' dimension described by \citet{Tarter2010}, who focused
on ``complex ... broadband signals''.  It also appears to be distinct
from the ``modulation'' dimension of \citet{wright18haystack}, who
contemplated drift rates on the order of the ``Earth's barycentric
acceleration'', i.e., 0.03~\Hzs at the center frequency of our
observations.
It is also absent from the CWTFM used by \citet{Enriquez2017,Price2020,wlod20}. 
The development of an improved figure of merit for
radio technosignature searches is beyond the scope of this work.
However, we recommend that improved figures of merit include the range
of
line-of-sight accelerations between emitter and receiver as a
dimension of the search volume as well as explicit guidelines
regarding the treatment of quantization and dechirping efficiencies.

\subsection{Re-analysis of 2016 and 2017 Data} \label{sec:1617Results}

\citet{Margot2018} presented the results of a search for
technosignatures around 14 planetary systems in the Kepler field
conducted on April 15, 2016, 16:00 - 18:00 universal time (UT) with
the GBT.  \citet{Pinchuk2019} presented the results of a similar
search conducted on May 4, 2017, 15:00 - 17:00 universal time (UT)
that included 10 planetary systems in the Kepler field but also
included scans of TRAPPIST-1 and LHS 1140.

We reprocessed these data with our updated algorithms and detected a
total of 13,750,469 candidate signals over the 2016 and 2017
epochs of observation.  Tables of signal properties of the detected
candidates are available online for both the 2016
\citep{seti16dataset}
and 2017 \citep{seti17dataset} data sets.
We found that 13,696,445 ($99.61\%$) signals were automatically
flagged as anthropogenic RFI and 54,024 signals were labeled as
promising.  Candidate signals found within operating regions of known
interferers (Table \ref{tab:known_RFI}) were attributed to RFI and
removed from consideration.  Visual inspection of all of the remaining
4,257 candidate signals revealed that they are attributable to RFI.
With this improved analysis, we confirm the initial results that no
technosignatures were detected in the data obtained in 2016
\citep{Margot2018} and 2017 \citep{Pinchuk2019}.

\section{Conclusions}\label{sec:conclusions}

We described the results of a search for technosignatures that used
4 hours of GBT time in 2018 and 2019. We identified \totcount candidate
signals, \filterremovedpercent$ $ of which were automatically
classified as RFI by rejection filters.  Of the signals that remained,
\finalcount were found outside of known RFI frequency bands and were
visually inspected.  All of these were attributable to RFI and none
were identified as a technosignature.

We presented significant improvements to our signal detection and
direction-of-origin filter algorithms.  We tested the signal recovery
of the updated procedures with a preliminary signal injection and
recovery analysis, which showed that our pipeline detects $\sim93\%$
of the injected signals overall.  This recovery rate increases to
$\sim98\%$ outside of known RFI frequency bands.  In addition, our
pipeline correctly identified \recoveredareFANDYpercent$ $ of the
artificial signals as technosignatures.
This signal injection and recovery analysis provides 
an important tool for quantifying the signal recovery rate of a
radio technosignature data processing pipeline.
Planned improvements to this tool will further illuminate
imperfections in our and other groups' pipelines and point to additional
areas for improvement.

Our search represents only a modest fraction of the BL searches
described by \citet{Enriquez2017} and \citet{Price2020} in terms of
number of targets and data volume.  However, our search strategy
has advantages
compared to these searches in terms of sensitivity
(up to 25 times better), frequency drift rate coverage (2--4 times
larger), and signal detection count per unit bandwidth per unit
integration time
($\sim$\replaced{300}{200} times larger).

We described limitations of recent Drake Figure of Merit (DFM)
calculations in assessing the probability of success of different
search programs.  These calculations have ignored important factors
such as quantization and dechirping efficiencies.  In addition, the
DFM does not take account of the range of drift rates considered in a
search nor the quality of signal detection algorithms.  As a result,
we suggest that recent DFM calculations are questionable indicators of
actual search volume or performance.
We recommend that improved metrics include the range of line-of-sight
accelerations between emitter and receiver as a dimension of the
search volume as well as explicit guidelines regarding the treatment
of quantization and dechirping efficiencies.

Our observations were designed, obtained, and analyzed by
undergraduate and graduate students enrolled in an annual SETI course
offered at UCLA since 2016.  The search for technosignatures can be
used effectively to teach skills in radio astronomy,
telecommunications, programming, signal processing, and statistical
analysis.  Additional information about the course is available at
\url{https://seti.ucla.edu}.

\acknowledgments

Funding for the UCLA SETI Group was provided by The Queens Road
Foundation, Janet Marott, Michael W. Thacher and Rhonda L. Rundle,
Larry Lesyna, and others donors (\url{https://seti.ucla.edu}).  Funding
for this search was provided by Michael W. Thacher and Rhonda
L. Rundle, Howard and Astrid Preston, K. K., Larry Lesyna, Herbert
Slavin, Robert Schneider, James Zidell, Joseph and Jennifer Lazio, and
25 other donors (\url{https://spark.ucla.edu/project/13255/wall}).
\added{We are grateful to a reviewer for useful suggestions.}
We are grateful to the BL team for stimulating discussions about
dechirping efficiency, data requantization, and data archival
practices.  We are grateful for the data processing pipeline initially
developed by the 2016 and 2017 UCLA SETI classes.  We thank Smadar
Gilboa, Marek Grzeskowiak, and Max Kopelevich for providing an
excellent computing environment in the Orville L. Chapman Science
Learning Center at UCLA.  We are grateful to Wolfgang Baudler, Paul
Demorest, John Ford, Frank Ghigo, Ron Maddalena, Toney Minter, and
Karen O’Neil for enabling the GBT observations.  The Green Bank
Observatory is a facility of the National Science Foundation operated
under cooperative agreement by Associated Universities, Inc. This work
has made use of data from the European Space Agency (ESA) mission {\it
  Gaia} (\url{https://www.cosmos.esa.int/gaia}), processed by the {\it
  Gaia} Data Processing and Analysis Consortium (DPAC,
\url{https://www.cosmos.esa.int/web/gaia/dpac/consortium}). This
research has made use of the SIMBAD database, operated at CDS,
Strasbourg, France.

\vspace{5mm}
\facilities{Green Bank Telescope}

\bibliography{seti18}

\begin{thebibliography}{}
\expandafter\ifx\csname natexlab\endcsname\relax\def\natexlab#1{#1}\fi
\providecommand{\url}[1]{\href{#1}{#1}}
\providecommand{\dodoi}[1]{doi:~\href{http://doi.org/#1}{\nolinkurl{#1}}}
\providecommand{\doeprint}[1]{\href{http://ascl.net/#1}{\nolinkurl{http://ascl.net/#1}}}
\providecommand{\doarXiv}[1]{\href{https://arxiv.org/abs/#1}{\nolinkurl{https://arxiv.org/abs/#1}}}

\bibitem[{{Christiansen} {et~al.}(2013){Christiansen}, {Clarke}, {Burke},
  {Jenkins}, {Barclay}, {Ford}, {Haas}, {Sabale}, {Seader}, {Claiborne Smith},
  {Tenenbaum}, {Twicken}, {Kamal Uddin}, \& {Thompson}}]{Christiansen2013}
{Christiansen}, J.~L., {Clarke}, B.~D., {Burke}, C.~J., {et~al.} 2013, \apjs,
  207, 35, \dodoi{10.1088/0067-0049/207/2/35}

\bibitem[{{Drake}(1984)}]{Drake1984}
{Drake}, F. 1984, {SETI Science Working Group Report}, ed. F.~{Drake}, J.~H.
  Wolfe, \& C.~L. Seeger, NASA Technical Paper No. TP-2244, 67--69

\bibitem[{{DuPlain} {et~al.}(2008){DuPlain}, {Ransom}, {Demorest}, {Brandt},
  {Ford}, \& {Shelton}}]{GUPPI}
{DuPlain}, R., {Ransom}, S., {Demorest}, P., {et~al.} 2008, in \procspie, Vol.
  7019, Advanced Software and Control for Astronomy II, 70191D,
  \dodoi{10.1117/12.790003}

\bibitem[{{Enriquez} {et~al.}(2017){Enriquez}, {Siemion}, {Foster}, {Gajjar},
  {Hellbourg}, {Hickish}, {Isaacson}, {Price}, {Croft}, {DeBoer}, {Lebofsky},
  {MacMahon}, \& {Werthimer}}]{Enriquez2017}
{Enriquez}, J.~E., {Siemion}, A., {Foster}, G., {et~al.} 2017, \apj, 849, 104,
  \dodoi{10.3847/1538-4357/aa8d1b}

\bibitem[{{Gaia Collaboration}(2016)}]{gaiadr1}
{Gaia Collaboration}. 2016, \aap, 595, A1, \dodoi{10.1051/0004-6361/201629272}

\bibitem[{{Gaia Collaboration}(2018)}]{gaiadr2}
---. 2018, \aap, 616, A1, \dodoi{10.1051/0004-6361/201833051}

\bibitem[{Gale \& Shapley(1962)}]{Gale1962}
Gale, D., \& Shapley, L.~S. 1962, The American Mathematical Monthly, 69, 9

\bibitem[{{Gray} \& {Mooley}(2017)}]{Gray2017}
{Gray}, R.~H., \& {Mooley}, K. 2017, \aj, 153, 110,
  \dodoi{10.3847/1538-3881/153/3/110}

\bibitem[{{Harp} {et~al.}(2016){Harp}, {Richards}, {Tarter}, {Dreher},
  {Jordan}, {Shostak}, {Smolek}, {Kilsdonk}, {Wilcox}, {Wimberly}, {Ross},
  {Barott}, {Ackermann}, \& {Blair}}]{Harp2016}
{Harp}, G.~R., {Richards}, J., {Tarter}, J.~C., {et~al.} 2016, \aj, 152, 181,
  \dodoi{10.3847/0004-6256/152/6/181}

\bibitem[{{Horowitz} \& {Sagan}(1993)}]{horo93}
{Horowitz}, P., \& {Sagan}, C. 1993, \apj, 415, 218, \dodoi{10.1086/173157}

\bibitem[{Kogan(1998)}]{Kogan1998}
Kogan, L. 1998, Radio Science, 33, 1289, \dodoi{10.1029/98RS02202}

\bibitem[{{Korpela}(2012)}]{korp12}
{Korpela}, E.~J. 2012, Annual Review of Earth and Planetary Sciences, 40, 69,
  \dodoi{10.1146/annurev-earth-040809-152348}

\bibitem[{{Lebofsky} {et~al.}(2019){Lebofsky}, {Croft}, {Siemion}, {Price},
  {Enriquez}, {Isaacson}, {MacMahon}, {Anderson}, {Brzycki}, {Cobb}, {Czech},
  {DeBoer}, {DeMarines}, {Drew}, {Foster}, {Gajjar}, {Gizani}, {Hellbourg},
  {Korpela}, {Lacki}, {Sheikh}, {Werthimer}, {Worden}, {Yu}, \&
  {Zhang}}]{lebo19}
{Lebofsky}, M., {Croft}, S., {Siemion}, A. P.~V., {et~al.} 2019, \pasp, 131,
  124505, \dodoi{10.1088/1538-3873/ab3e82}

\bibitem[{Margot {et~al.}(2020{\natexlab{a}})Margot, Greenberg, Pinchuk,
  {et~al.}}]{seti16dataset}
Margot, J.~L., Greenberg, A.~H., Pinchuk, P., {et~al.} 2020{\natexlab{a}}, Data
  from: A search for technosignatures from 14 planetary systems in the Kepler
  field with the Green Bank Telescope at 1.15–1.73 GHz, v4, Dataset,
  \dodoi{10.5068/D1309D}

\bibitem[{Margot {et~al.}(2020{\natexlab{b}})Margot, Pinchuk, Greenberg,
  {et~al.}}]{seti17dataset}
Margot, J.~L., Pinchuk, P., Greenberg, A.~H., {et~al.} 2020{\natexlab{b}}, Data
  from: A search for technosignatures from TRAPPIST-1, LHS 1140, and 10
  planetary systems in the Kepler field with the Green Bank Telescope at
  1.15–1.73 GHz, v4, Dataset, \dodoi{10.5068/D1Z964}

\bibitem[{{Margot} {et~al.}(2018){Margot}, {Greenberg}, {Pinchuk}, {Shinde},
  {Alladi}, {Prasad MN}, {Bowman}, {Fisher}, {Gyalay}, {McKibbin}, {Miles},
  {Nguyen}, {Power}, {Ramani}, {Raviprasad}, {Santana}, \&
  {Lynch}}]{Margot2018}
{Margot}, J.-L., {Greenberg}, A.~H., {Pinchuk}, P., {et~al.} 2018, \aj, 155,
  209, \dodoi{10.3847/1538-3881/aabb03}

\bibitem[{{Pinchuk} {et~al.}(2019){Pinchuk}, {Margot}, {Greenberg}, {Ayalde},
  {Bloxham}, {Boddu}, {Gerardo Chinchilla-Garcia}, {Cliffe}, {Gallagher},
  {Hart}, {Hesford}, {Mizrahi}, {Pike}, {Rodger}, {Sayki}, {Schneck}, {Tan},
  {{\textquotedblleft}Yolanda{\textquotedblright} Xiao}, \&
  {Lynch}}]{Pinchuk2019}
{Pinchuk}, P., {Margot}, J.-L., {Greenberg}, A.~H., {et~al.} 2019, \aj, 157,
  122, \dodoi{10.3847/1538-3881/ab0105}

\bibitem[{{Price} {et~al.}(2020){Price}, {Enriquez}, {Brzycki}, {Croft},
  {Czech}, {DeBoer}, {DeMarines}, {Foster}, {Gajjar}, {Gizani}, {Hellbourg},
  {Isaacson}, {Lacki}, {Lebofsky}, {MacMahon}, {Pater}, {Siemion}, {Werthimer},
  {Green}, {Kaczmarek}, {Maddalena}, {Mader}, {Drew}, \& {Worden}}]{Price2020}
{Price}, D.~C., {Enriquez}, J.~E., {Brzycki}, B., {et~al.} 2020, \aj, 159, 86,
  \dodoi{10.3847/1538-3881/ab65f1}

\bibitem[{{Sheikh} {et~al.}(2019){Sheikh}, {Wright}, {Siemion}, \&
  {Enriquez}}]{Sheikh2019}
{Sheikh}, S.~Z., {Wright}, J.~T., {Siemion}, A., \& {Enriquez}, J.~E. 2019,
  \apj, 884, 14, \dodoi{10.3847/1538-4357/ab3fa8}

\bibitem[{{Siemion} {et~al.}(2013){Siemion}, {Demorest}, {Korpela},
  {Maddalena}, {Werthimer}, {Cobb}, {Howard}, {Langston}, {Lebofsky}, {Marcy},
  \& {Tarter}}]{Siemion2013}
{Siemion}, A.~P.~V., {Demorest}, P., {Korpela}, E., {et~al.} 2013, \apj, 767,
  94, \dodoi{10.1088/0004-637X/767/1/94}

\bibitem[{{Tarter}(2001)}]{Tarter2001}
{Tarter}, J. 2001, Annual Review of Astronomy and Astrophysics, 39, 511,
  \dodoi{10.1146/annurev.astro.39.1.511}

\bibitem[{{Tarter} {et~al.}(2010){Tarter}, {Agrawal}, {Ackermann}, {Backus},
  {Blair}, {Bradford}, {Harp}, {Jordan}, {Kilsdonk}, {Smolek}, {Richards},
  {Ross}, {Shostak}, \& {Vakoch}}]{Tarter2010}
{Tarter}, J.~C., {Agrawal}, A., {Ackermann}, R., {et~al.} 2010, in Society of
  Photo-Optical Instrumentation Engineers (SPIE) Conference Series, Vol. 7819,
  Instruments, Methods, and Missions for Astrobiology XIII, 781902,
  \dodoi{10.1117/12.863128}

\bibitem[{{Taylor}(1974)}]{Taylor1974}
{Taylor}, J.~H. 1974, \aaps, 15, 367

\bibitem[{{Virtanen} {et~al.}(2020){Virtanen}, {Gommers}, {Oliphant},
  {Haberland}, {Reddy}, {Cournapeau}, {Burovski}, {Peterson}, {Weckesser},
  {Bright}, {van der Walt}, {Brett}, {Wilson}, {Jarrod Millman}, {Mayorov},
  {Nelson}, {Jones}, {Kern}, {Larson}, {Carey}, {Polat}, {Feng}, {Moore}, {Vand
  erPlas}, {Laxalde}, {Perktold}, {Cimrman}, {Henriksen}, {Quintero}, {Harris},
  {Archibald}, {Ribeiro}, {Pedregosa}, {van Mulbregt}, \&
  {Contributors}}]{SciPy}
{Virtanen}, P., {Gommers}, R., {Oliphant}, T.~E., {et~al.} 2020, Nature
  Methods, 17, 261, \dodoi{https://doi.org/10.1038/s41592-019-0686-2}

\bibitem[{{Wenger} {et~al.}(2000){Wenger}, {Ochsenbein}, {Egret}, {Dubois},
  {Bonnarel}, {Borde}, {Genova}, {Jasniewicz}, {Lalo{\"e}}, {Lesteven}, \&
  {Monier}}]{Wenger2000}
{Wenger}, M., {Ochsenbein}, F., {Egret}, D., {et~al.} 2000, \aaps, 143, 9,
  \dodoi{10.1051/aas:2000332}

\bibitem[{{Wlodarczyk-Sroka} {et~al.}(2020){Wlodarczyk-Sroka}, {Garrett}, \&
  {Siemion}}]{wlod20}
{Wlodarczyk-Sroka}, B.~S., {Garrett}, M.~A., \& {Siemion}, A.~P.~V. 2020, arXiv
  e-prints, arXiv:2006.09756.
\newblock \doarXiv{2006.09756}

\bibitem[{{Wright} {et~al.}(2018){Wright}, {Kanodia}, \&
  {Lubar}}]{wright18haystack}
{Wright}, J.~T., {Kanodia}, S., \& {Lubar}, E. 2018, Astronomical Journal, 156,
  260, \dodoi{10.3847/1538-3881/aae099}

\end{thebibliography}

\end{document}